\documentclass[%
 reprint, 
 aps,pre,
 twocolumn,
 10pt,
 showkeys,groupaddress,preprintnumbers
 ]{revtex4-2}

\usepackage[english]{babel}
\usepackage{graphicx}
\usepackage{dcolumn}
\usepackage{bm}
\usepackage{siunitx}  
\usepackage{amsmath}
\usepackage{tabularx}

\usepackage{subfigure}

\usepackage[normalem]{ulem}
\usepackage{blkarray}
\usepackage[hidelinks]{hyperref}
\usepackage{cancel}
\usepackage{tikz}

\usepackage{array}
\newcolumntype{P}[1]{>{\centering\arraybackslash}p{#1}}

\newcommand{\p}[1][]{\varphi_{#1}}
\newcommand{\pdc}[1][]{\varphi_{{#1}dc}}
\newcommand{\pxdc}[1][]{\varphi_{{#1}xdc}}
\newcommand{\px}[1][]{\varphi_{{#1}x}}

\begin{document}

\topmargin=-0.45in
\evensidemargin=0in
\oddsidemargin=0in
\textwidth=6.5in

\textheight=9.0in
\headsep=0.25in 

\title{Momentum-Driven Reversible Logic\\
Accelerates\\
Efficient Irreversible Universal Computation}

\author{Kuen Wai Tang}
\email{tkwtang@ucdavis.edu}

\author{Kyle J. Ray}
\email{kjray@ucdavis.edu}

\author{James P. Crutchfield}
\email{chaos@ucdavis.edu}
\affiliation{Department of Physics and Astronomy\\
University of California-Davis\\
One Shields Avenue, Davis, CA 95616}

\date{\today}

\preprint{arxiv.org:2602.XXXXX}

\begin{abstract}
We present implementations of two physically-embedded computation-universal logical operations using a 2-bit logical unit composed of coupled quantum flux parametrons—Josephson-junction superconducting circuits. To illustrate universality, we investigate NAND gates built from these two distinct elementary operations. On the one hand, Controlled Erasure (CE) is designed using fixed-point analysis and assumes that information must be stored in locally-metastable distributions. On the other, Erasure-Flip (EF) leverages momentum as a computational resource and significantly outperforms the metastable approach, simultaneously achieving higher fidelity and faster computational speed without incurring any additional energetic cost. Notably, the momentum degree of freedom allows the EF to achieve universality by using both nontrivial reversible and irreversible logic simultaneously in different logical subspaces. These results not only provide a practical, high-performance protocol ripe for experimental realization but also underscore the broader potential of momentum-based computing paradigms. 
\end{abstract}


\maketitle

\section{Introduction}
\label{sec:Introduction}

The information data industry is currently facing an unprecedented energy demand crisis. This challenge is driven by the rapid growth of data centers and the proliferation of digital services \cite{bogmans2025power, chen2025electricity, masanet2020recalibrating}. Historically, the industry's ability to consistently meet the growth in computational demand  has been driven by the scaling achieved under Moore's Law \cite{moore1964future, moore1998cramming}. However, as conventional CMOS technology approaches fundamental physical limits during scaling-down  \cite{wong2005road, khan2021nanoscale, radamson2024cmos, kuroda2001cmos, frank2005limits}, particularly regarding thermal dissipation, the need for sustainable, energy-efficient hardware is more critical than ever. Addressing this global energy cost problem requires the development of novel computing design paradigms capable of operating at or near the fundamental thermodynamic limits of computation.

One such effort targets low-energy circuits based on the quantum flux parametron (QFP), a superconducting logical element that operates in the classical regime \cite{1063734, PhysRevLett.75.1614, variable_beta, likharev1982classical, 1487397, hosoya1991quantum, hioe1991quantum} rather than the quantum regime. Various computational architectures based on QFPs have been proposed. For example, the single-flux quantum (SQF) \cite{1059351} and rapid single-flux quantum (RSQF) \cite{1064951, 783712, tanaka201218} use voltage pulses to process information. Adiabatic QFPs \cite{takeuchi2022adiabatic, 10.1063/1.5080753, PhysRevApplied.4.034007, 10.1063/1.4790276, chen2019adiabatic} and reversible QFPs \cite{takeuchi2014reversible, takeuchi2017reversibility, yamae2024minimum} store information in the  components of a QFP's phase degrees of freedom and logic gate protocols are then designed assuming adiabatic operation---which assumes fast and reliable relaxation to a local steady-state relative to the protocol's timescale. Furthermore, other interesting reversible logic concepts utilizing superconducting circuits have been extensively studied \cite{semenov2003negative, semenov2007classical, osborn2019reversible, frank2019asynchronous, wustmann2020reversible}. 

In many operating regimes, the phase-space evolution of these devices are well modeled by underdamped Langevin dynamics \cite{kalmykov2012langevin}. This permits transient dynamics that can be dominated by the conjugate momentum coordinate in phase space instead of just the position. The attempt to harness these transient dynamics for useful computation provides a compelling alternative design principle for logic gates that we call \emph{momentum computing} \cite{PhysRevResearch.3.023164}. Single-bit momentum computing has been demonstrated using a feedback driven electromechanical cantilever \cite{dago2023logical}, and a design has been proposed for an implementation based on a QFP-like circuit \cite{ray2022gigahertz}. 

Coupling two QFP's inductively creates a device capable of operations on two logical bits instead of one. The following investigates the thermodynamics of universal computation via control protocols that implement a NAND operation using such a coupled quantum flux parametron (CQFP) device. Given that NAND is a universal logic gate, energy-efficient implementations position CQFPs as compelling candidates for next-generation, high-speed, low-power computational devices.

To this end, the following analyzes and compares two distinct low-power protocols, each implementing NAND logic. First, we consider the Controlled Erasure (CE) protocol \cite{pratt2024}, which is designed using the principle of storing and processing information only in positional degrees of freedom. Next, we present a new universal momentum computing logic gate, the Erasure-Flip (EF). Remarkably, by leveraging the transient dynamics that conjugate momenta allow, the EF gate offers a NAND implementation that significantly increases both computational speed and fidelity without incurring additional energetic cost. 

Section \ref{section:CQFP} presents the CQFP circuit design and details the underlying principles for manipulating its potential landscape to perform logical computations. Section \ref{section:NAND_operation_intro} introduces the idea of implementing NAND operations in CQFPs by serially implementing more fundamental operations. Section \ref{section: controlled_eraure} then provides an in-depth analysis of the work cost, fidelity, and speed of the CE protocol, highlighting its inherent speed limitations. Addressing this, Section \ref{section:erasure_flip_protocol} introduces a novel control scheme---the EF protocol---that leverages the system momentum to preserve information. We demonstrate through simulations that this protocol achieves significantly higher speed and fidelity than the CE while maintaining low work cost. Finally, Section \ref{section:NAND_protocol} shows the simulation results of using both the CE and EF subprotocols introduced in the preceding sections to perform the NAND operation. The development concludes in Section \ref{section:conclusion} with a summary of our findings.

\section{Results}
\subsection{Coupled Quantum Flux Parametron}

\label{section:CQFP}

\subsubsection{Circuit}

\begin{figure}[ht]
\includegraphics[scale=0.5]{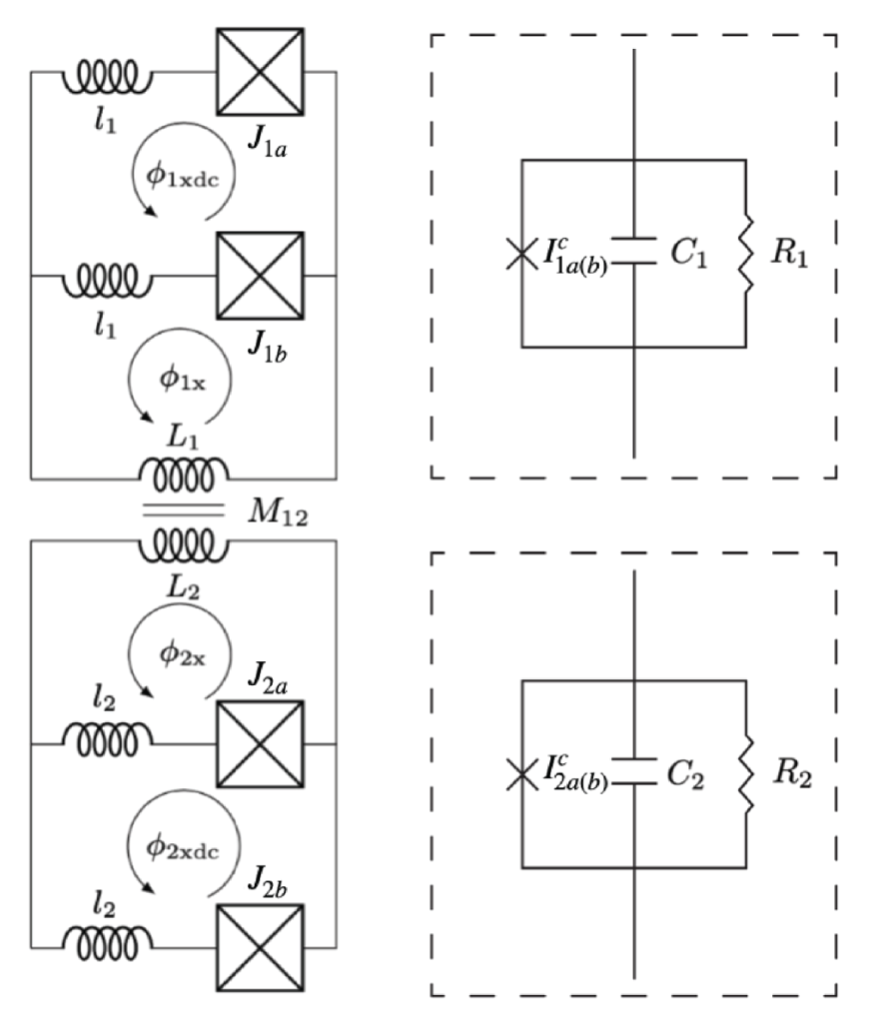}
\caption{CQFP Circuit: Boxed diagrams (dashed lines) represent the Josephson junctions modeled as RCSJs. Table \ref{table:circuit_parameters} in the Appendix explains the symbols.}
\label{fig:cqfp_circuit_diagram.png}
\end{figure}

A CQFP is formed by inductively coupling two QFPs \cite{PhysRevApplied.13.034037, vanDenBrink_2005, PhysRevLett.98.177001, PhysRevB.80.052506}. The coupling mechanism is often a third QFP, whose dynamics we represent with a time-dependent coupling parameter $M_{12}$. Figure \ref{fig:cqfp_circuit_diagram.png} shows the CQFP circuit diagram. The Josephson junctions of the $i$-th QFP are denoted $J_{ia}$ and $J_{ib}$ with phase differences $\delta_{ia}$ and $\delta_{ib}$, respectively. Each junction is modeled as a resistive capacitive shunted junction (RCSJ) \cite{1968ApPhL..12..277S, 10.1063/1.1656743}. $C_i$ and $R_i$ represent the junctions' shunt capacitors and resistors. $I^c_{ia}$ and $I^c_{ib}$ are the junctions critical currents. Together, $J_{ia}$ and $J_{ib}$ form a dc-SQUID with inductance $l_i$ in each branch. This dc-SQUID is then connected in a loop with a geometric inductance $L_i$, as in an rf-SQUID. For simplicity, we set $l_1 = l_2 = l$, $L_1 = L_2 = L$, $C_1 = C_2 = C$, and $R_1 = R_2 = R$. The device's characteristic timescale is  given by $t_c = \sqrt{LC}$ which, for the specific parameters we consider, is $2.24 \text{ps}$.

The device's state is described by a four-dimensional vector: $(\phi_1, \phi_2, \phi_{1dc}, \phi_{2dc})$. Here,
$\phi_1$ and $\phi_2$ represent the flux through the rf-SQUID of the two constituent QFPs, while $\phi_{1dc}$ and $\phi_{2dc}$ represent the flux through their small dc-SQUID loops. These dynamical coordinates can be expressed in terms of the JJ phases as:
\begin{align*}
 \phi_i &= \frac{\Phi_0}{2\pi}\frac{\delta_{ia} + \delta_{ib}}{2} \\
 \phi_{idc} &= \frac{\Phi_0}{2\pi}(\delta_{ia} - \delta_{ib})
 ~. 
\end{align*}
 
The shunt DC resistance $R$ acts as a source of noise and damping, which allows us to model the dynamics of these coordinates using an underdamped Langevin equation. Since the system's state evolves according to Langevin dynamics (see Appendix \ref{appendix:langevin_dynamics}), we use the term ``particle'' as an analogy to describe the device's state and extend this metaphor by using the terms ``position'' and ``velocity'' to refer to values of the coordinate and its time derivative. This is a conceptual representation to motivate an intuitive understanding and is not meant to imply there is an actual physical particle.

The driving force in the Langevin equation that describes the motion of these particles can be written as the derivative of a potential landscape manipulated using five external parameters. These include external fluxes applied to the large loops ($\phi_{ix}$) and the junction loops ($\phi_{ixdc}$), as well as an inductive coupling between the two QFPs ($M_{12}$). The corresponding normalized external parameters are $\px[i]$, $\pxdc[i]$, and $m_{12}$ respectively. The relationships are given by:
\begin{equation*}
    \px[i] = \frac{2\pi\phi_{ix}}{\Phi_0}, \pxdc[i] = \frac{2\pi\phi_{ix}}{\Phi_0}, ~\text{and}~ m_{12} = \frac{M_{12}}{\sqrt{L_1 L_2}},
\end{equation*}
where $\Phi_0$ is the magnetic flux quantum. The parameters introduced above and their corresponding values used in the simulations are summarized in Table \ref{table:circuit_parameters} in Appendix \ref{appendix:detail_of_the_simulation}.

\subsubsection{Potential}

The potential energy function associated with the single QFP is well known \cite{1487397, hioe1991quantum, variable_beta, takeuchi2022adiabatic}. From it, one can find the CQFP's potential energy \cite{PhysRevResearch.7.013014, Tang_2025}, which we write in the following nondimensionalized form:
\begin{align} 
\label{eq:QFP_potential}
U' = U_1 + U_2 + U_{12}
  ~,
\end{align}
with:
\begin{align*}
    U_i &= \frac{\xi}{2}(\p[i] - \px[i])^2 + \frac{\gamma_i}{2}(\pdc[i]-\pxdc[i])^2 \\
    & +\beta_i \cos\p[i] \cos\frac{\pdc[i]}{2} 
     + \delta \beta_i \sin\p[i] \sin\frac{\pdc[i]}{2} \\
    U_{12} &= m_{12} \xi(\p[1] - \px[1])(\p[2] - \px[2]),\\
\end{align*}
where the index i = 1, 2.

Here, $\xi$ is a modified coupling parameter $\xi \equiv 1/(1 - m_{12}^2)$ with $\xi \approx 1$ indicating the weak coupling limit. The parameter $\gamma_i$ represents the ratio between the geometric inductances of the rf-SQUID loop and the dc-SQUID loop and is typically set to be $>>1$. Due to this large ratio, any significant deviation of $\pdc[i]$ from $\pxdc[i]$ incurs a large energy penalty, especially when compared to the deviation of $\p[i]$ from $\p[ix]$. For this reason, it is often assumed that $\pdc[i] \approx \pxdc[i]$, allowing simplification of the potential to a 2D landscape in $\p[1]$ and $\p[2]$ space. The parameters $\beta_i$ and $\delta \beta_i$ are proportional to the sum and difference of the critical currents in the Josephson junctions, respectively. For simplification, the critical currents are assumed to be identical, resulting in $\delta \beta_i = 0$. However, in reality, it is unavoidable to have slightly asymmetric Josephson junctions, which produces an asymmetric potential that inherently favors some particular states. Fortunately, this asymmetric problem can be alleviated by calibration of the potential landscape; see Refs. \cite{ray2022gigahertz, Tang_2025}. The full expressions for $\gamma_i$, $\beta_i$ and $\delta \beta_i$ are provided in Table \ref{table:circuit_parameters}. 

When all external parameters are set to zero, the system exhibits a four-well potential within the $\p[1]$-$\p[2]$  state space, as visualized in the contour plot in Figure \ref{fig:Controlled_Erasure_protocol}(a). Each of these potential minima is capable of storing information as a metastable state, because the energy scale of these wells is significantly higher than the thermal energy scale $k_B T$ in typical operating regimes. The logical states are coarse-grained based on the sign of the individual phases: $\p[i] < 0$ represents logical $0$ and $\p[i] > 0$ represents logical $1$. Consequently, each well is labeled with its corresponding binary state: $00$, $01$, $10$, and $11$. The effects of the external control parameters on the system's potential energy landscape are further illustrated in Figure \ref{fig:parameter_functions} of Appendix \ref{appendix:detail_of_protocols}.

The colored dots in the contour plot at $t=0$ in Figure \ref{fig:Controlled_Erasure_protocol} represent samples from the initial equilibrium distribution. The coarse-grained initial position serves as the logical input for the operation, with color indicating the initial logical state of each sample. By dynamically manipulating the potential landscape, the particles are guided through the state space, representing the evolution of the system's microstates. The logical output of the operation is determined by a coarse-grained measurement of the final positions of the particles.

\subsection{Physically-Embedded NAND}
\label{section:NAND_operation_intro}



\begin{figure}[ht]
\includegraphics[scale=0.7]{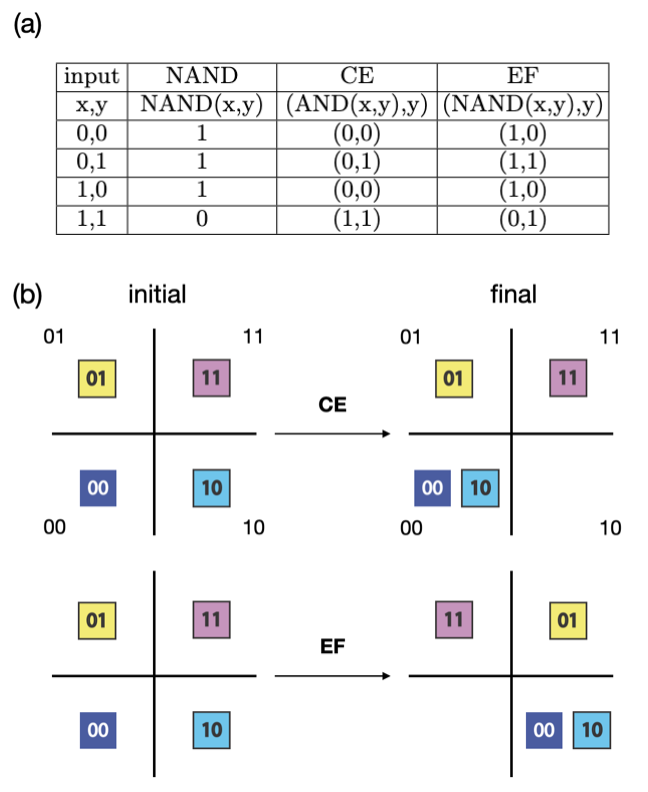}
\caption{(a) Truth tables for the three operations: NAND, Controlled Erasure (CE), and Erasure-Flip (EF). (b) Schematic illustration of the initial and final logical-state distributions for CE (top) and EF (bottom) operations. The four 2-bit digits at the corners indicate the coarse-graining label for the potential wells and the colored boxes represent the particles. The initial state (left) represents the thermal equilibrium distribution of particles within the four-well potential, corresponding to the logical inputs. The final state (right) depicts the particle distribution after the respective operations, which should be compared with the truth table to determine the logical output.}
\label{fig:truth_table_and_three_operations}
\end{figure}

The NAND operation is a universal two-input logic gate whose output is zero if and only if both inputs are one; otherwise, the output is one. NAND gates are a favored logical element in modern  general-purpose microprocessor architecture, and so the ability to simulate a NAND is a common standard candle for determining an operation's computation universality. Figure \ref{fig:truth_table_and_three_operations}(a) gives the NAND truth table. Although the NAND operation is a $2$-to-$1$ mapping, the CQFP produces a two-bit output. And, in fact, this will be the case for any physically embedded cyclic computation that stores $2$-bit information in $2$ degrees of freedom. If we call the input bits $x$ ane $y$ and the output bits $x'$ and $y'$, then an arbitrary operation is given by $ (x',y') = (f(x,y),g(x,y))$ where $f(\cdot)$ and $g(\cdot)$ are functions of the input memory states. 

A NAND gate, in this context, is an underdetermined concept. It could be the case that only $f(x,y)=\text{NAND}(x,y)$ or only $g(x,y)=\text{NAND}(x,y)$. Thus, when dealing with such physically embedded logic gates, it is more accurate to say that an operation on the memory states of the device can ``contain" or ``implement'' a NAND gate rather than ``be" a NAND gate. We address this ambiguity by defining a ``partial" NAND operation, which selects one of the two output bits as the result and is agnostic to the other one. (This terminology is common to the theory of computation \cite{Lewi98a}.) 

The EF operation in Figure \ref{fig:truth_table_and_three_operations}(b) illustrates the memory state mapping in the CQFP state space as an example of a partial NAND. In this scenario, the initial equilibrium distributions representing inputs 00, 01, and 10 are mapped to a final logical outputs of either 10 or 11, while the input state 11 is mapped to 01. This partial NAND is described by $ (x',y') = (\text{NAND}(x,y),\text{y})$. That being said, if a NAND gate is the only desired logic then we consider the second bit simply as an irrelevant bit, while the first is assumed to be the output.

A second approach---the ``complete" NAND---removes ambiguity by requiring $f=g=\text{NAND}(x,y)$. While there is no ambiguity in the map choice for this gate, our focus here is the partial NAND due to the flexibility it affords. Additional details regarding implementations of complete NAND operations are provided in Appendix \ref{appendix:alternative_NAND}.

The absolute minimum possible work required for an operation is determined by the change in the system's entropy. Since we assume a cyclic protocol and the initial wells have identical local free energies, the change in the system's average internal energy is zero after local equilibration.
According to the first law of thermodynamics $\Delta U = W + Q = 0$. That is, the work performed ($W$) is the negative of the heat dissipated ($Q$).

For a reversible, isothermal process, $Q$ is given by the change in system entropy. This yields a minimal work value $W = -k_BT \Delta S$, which is often called Landauer's bound \cite{Landauer_1961}. The fundamental thermodynamic cost of the partial NAND gate is $0.347~k_BT$ (Appendix \ref{appendix:fundamental_work_cost}). With this in mind, we note that the finite-time protocols we investigate fall typically between one and two orders of magnitude higher in work cost than this limit.   


\subsection{Mechanism of Controlled Erasure}
\label{section: controlled_eraure}

This section describes the Controlled Erasure (CE) protocol, first proposed in Ref. \cite{pratt2024}, and, using extensive simulations, analyzes its work cost, fidelity, and speed. The key information processing of the CE protocol's is to erase one pair of inputs while keeping the other pair distinct. The truth table for one CE operation, along with the initial and final particle distributions, is illustrated in Figure \ref{fig:truth_table_and_three_operations}. Under this protocol, both the $00$ and $10$ particles are mapped to the $00$ well, while the $01$ and $11$ particles remain in their initial wells. Therefore, the protocol performs a conditional erasure: the first bit is erased to $0$ if and only if the second bit is $0$.

We refer to the $00$ and $10$ pair as the erasure pair and the $01$ and $11$ pair as the storage pair. In terms of binary logic, this CE returns $x'=\text{AND}(x,y)$ and $y'=y$. It is worth noting that we cannot think of CEs in general as returning an AND and an identity. Another choice of CE that does the same information processing erases $00$ and $10$ to $10$ (instead of $00$), resulting in $(x', y') = (\text{OR}(x,\text{NOT}(y)),y)$. The full set of logical operations that can be implemented using CE protocols are detailed in Ref. \cite{pratt2024}.

\begin{figure*}[ht]
\includegraphics[scale=0.45]{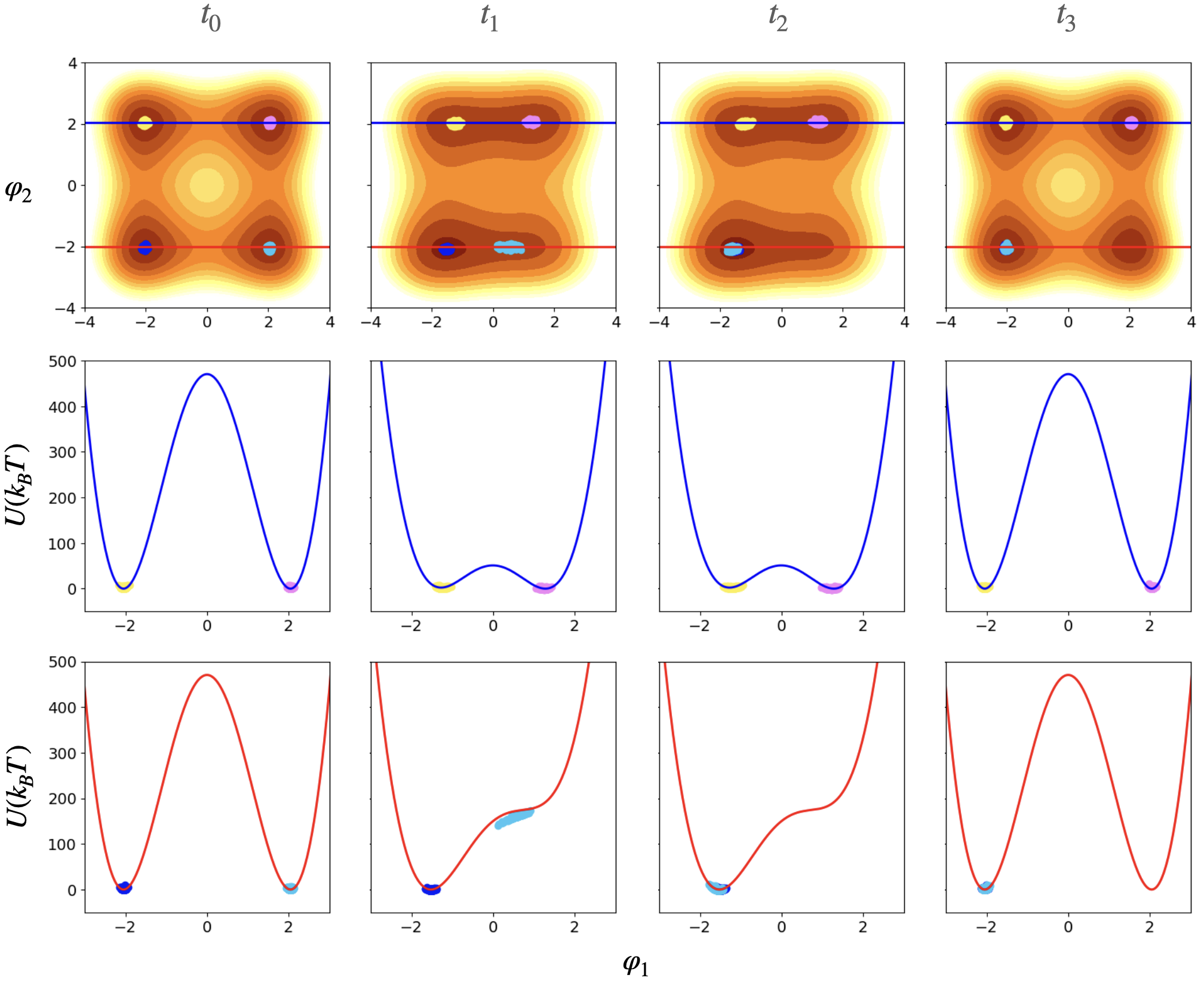}
\caption{Snapshots illustrating the evolution of the CE protocol are presented at four distinct time points. The first row of the figure contains the contour plots of the potential energy landscape in the $\p[1]\text{-}\p[2]$ state space, where the colored circles mark the instantaneous distributions of the four particle types (00, 01, 10, and 11). The second and third rows depict the corresponding potential energy graphs along the blue and red cutlines, respectively, as indicated on the contour plots. (\textcolor{blue}{\href{https://drive.google.com/file/d/1dn8LQAMVQNqLFT4am6F1jOUtWKFEmted/view?usp=sharing}{Animation available online}}).}
\label{fig:Controlled_Erasure_protocol}
\end{figure*}

\begin{figure*}
\includegraphics[scale=0.58]{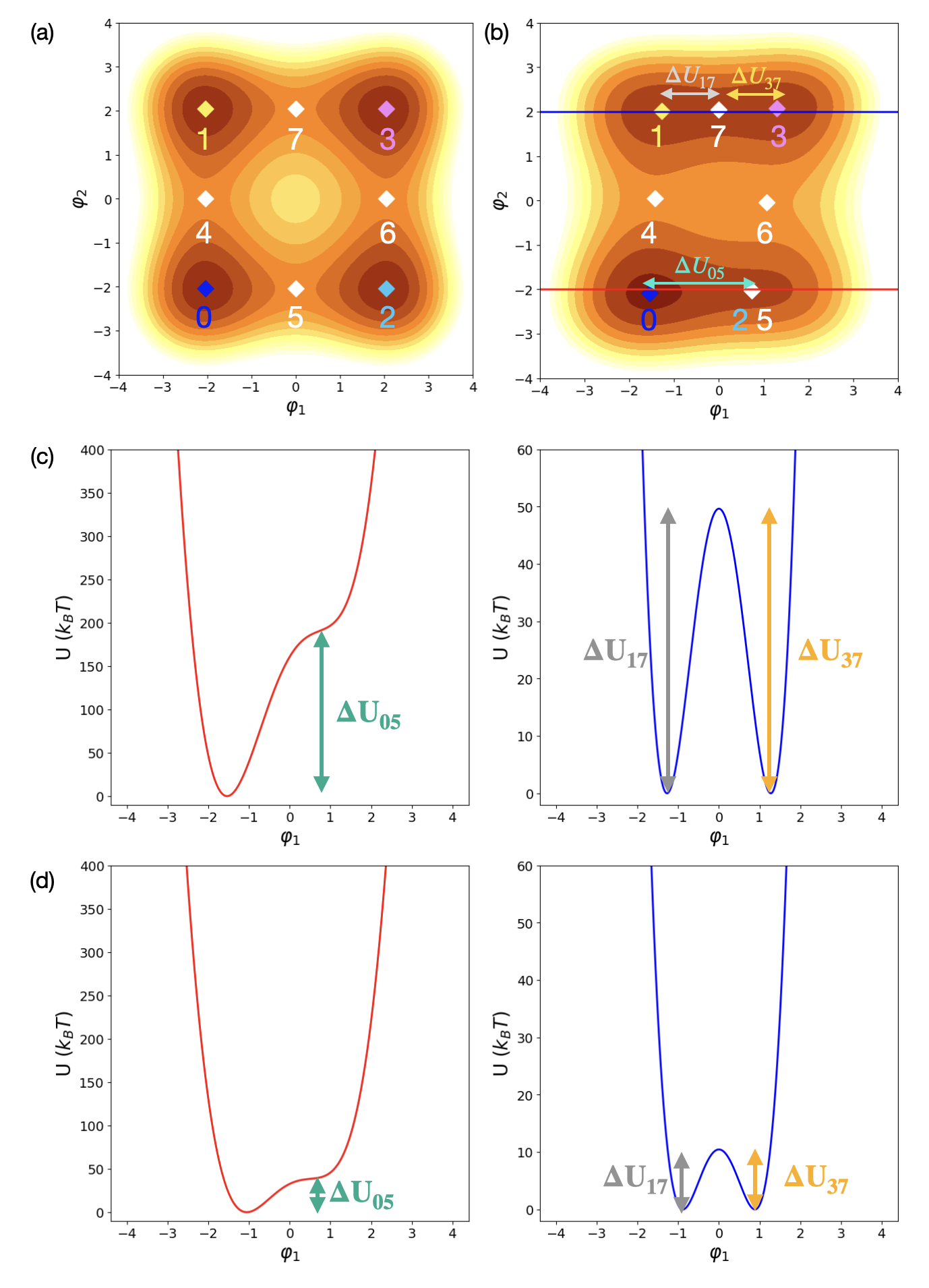}
\caption{(a) Contour graph of the four-well potential. The diamonds indicate critical points of the potential. Number 0 to 3 are the local minima and number 4 to 7 are saddle points.  (b) Potential contour at the moment of saddle point bifurcation happens. $\Delta U_{ij}$ indicate the potential differences between the critical points. The first index and second index in the subscript indicate the the local minimum and saddle point refering to. At the bifurcation  point, stable point 2 and saddle point 5 merge together. (c) The left (right) potential plots trace the potential along the red (blue) line in (b). (d) Reducing the potential barrier height $\Delta U_{05}$ leads to a decrease in the overall work cost of the CE protocol. These two graphs illustrate the potential energy profile along the respective cutlines when a significantly lower barrier height is employed.}
\label{fig:CE_potential_graphs}
\end{figure*}


Figure \ref{fig:Controlled_Erasure_protocol} displays snapshots of the potential landscape in the $\p[1]-\p[2]$ space at four distinct times in the CE cycle. The graphs below the contour plots illustrate the potential along the blue cutline (at $\p[2] = 2.3$) and the red cutline (at $\p[2] = -2.3$). The dots in the contour plots represent samples from the distribution of particles at that particular moment, with the color denoting the initial memory state of each sampled state.

The cycle begins at $t_0$ with a four-well potential. A detailed illustration of four-well potential is provided in Figure \ref{fig:CE_potential_graphs}(a), where the numbered diamonds identify the critical points in the potential's energy landscape: Points $0$ through $3$ are local minima and Points $4$ through $7$ are saddle points.

At $t_1$, the potential is manipulated to achieve a saddle-node bifurcation. Figure \ref{fig:CE_potential_graphs}(b) shows a detailed illustration of the potential at that particular time, showing how the critical points have evolved from $t_0$. The potential differences between these critical points are labeled as $\Delta U_{ab}$, where the indices of the subscript indicate the points to which they refer. Figure \ref{fig:CE_potential_graphs}(c) shows the potential along the blue and red cutlines. At this moment in the operation, the local minimum at Point $2$ merges with the saddle point at Point $5$.  The elimination of $\Delta U_{25}$ removes the barrier that previously separated the $00$ and $10$ regions. As a consequence, the $10$ particles (light blue) slide down a potential gradient defined by the height $\Delta U_{05}$, gaining kinetic energy. Concurrently, the potential barriers $\Delta U_{17}$ and $\Delta U_{37}$ are maintained at approximately $50~k_BT$. This barrier height is sufficient to ensure a negligible escape rate ($\approx 10^{-20}$ per operation) for the particles of the storage pair ($01$ and $11$) during the protocol duration, confining them to their initial wells.

The potential is held constant at this bifurcation point until $t_2$, allowing the $10$ particles to fully transition into the $00$ well. The kinetic energy gained during the descent is subsequently dissipated into the surrounding environment as heat, causing the particles to equilibrate. This process effectively erases the information that previously distinguished the $00$ and $10$ states. Throughout this process, the $01$ and $11$ particles remain isolated in their initial wells. This dissipative process dominates the protocol's energetic cost.

An alternative approach would use a pitchfork bifurcation, instead of a saddle-node bifurcation to merge the $00$ and $10$ wells. However, this is precluded by the necessity of maintaining high barriers to preserve the information stored in the $10$ and $11$ wells \cite{pratt2024}.

It is important to note that the motion of the $10$ particles during the CE protocol is predominantly in the $\p[1]$ direction, as the barriers separating the top and bottom wells are high enough (about $480~k_BT$) to prevent movement over the barrier. This allows their motion to be effectively modeled as one-dimensional. Ref. \cite{pratt2024} provides a range of circuit parameters that enable effective CE, establishing the device's robustness against logical errors.

Finally, from $t_2$ to $t_3$, the potential is restored to the initial four-well potential, completing one full cycle of the CE operation.

\subsection{Simulation Results of CE}

In this subsection, we present the simulation results for the CE protocol. The detailed equations and numerical methods used for these simulations are provided in Appendix \ref{appendix:detail_of_the_simulation}.

\subsubsection{Work Distributions}

\begin{figure}
\includegraphics[scale=0.5]{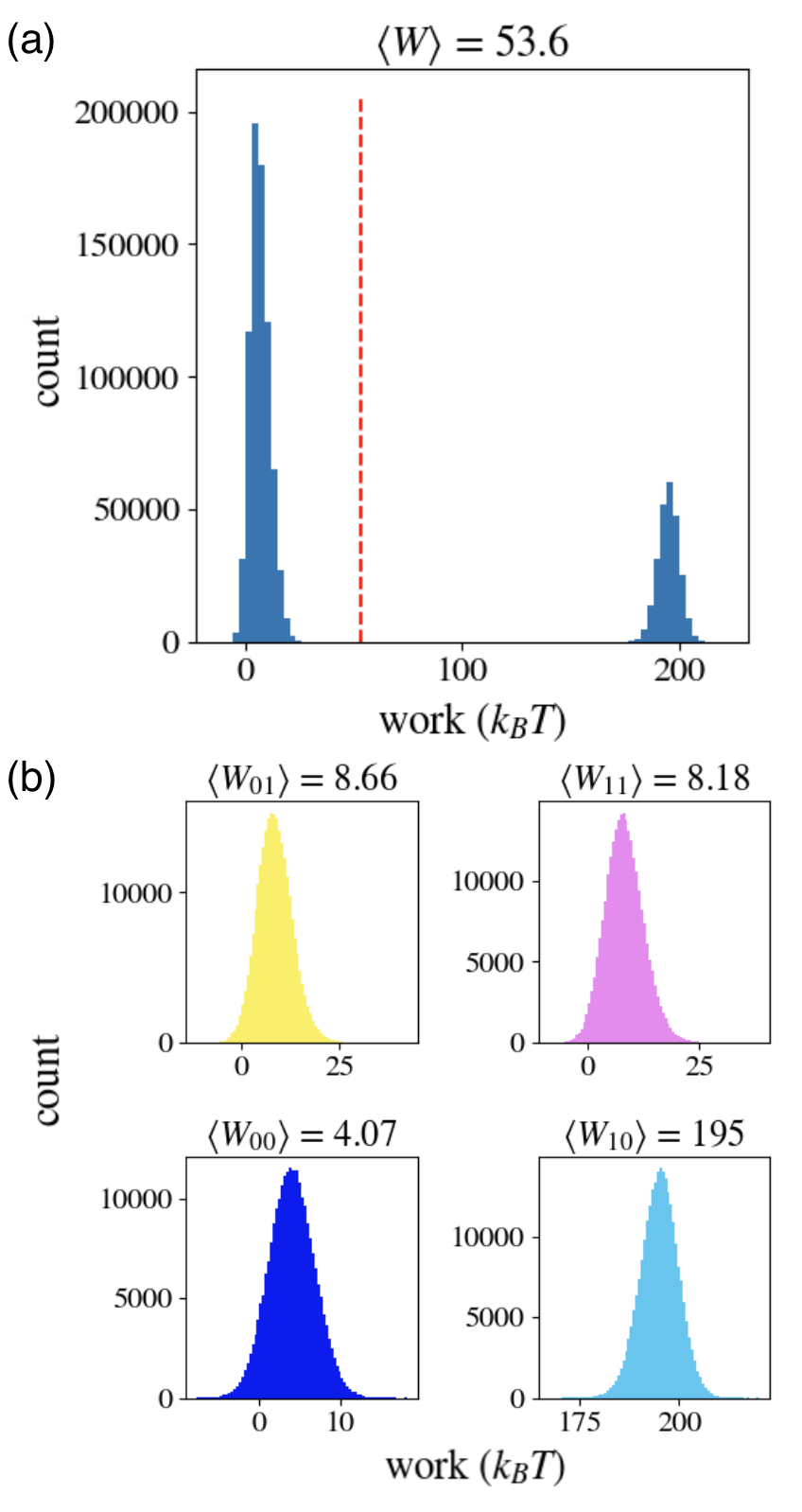}
\caption{(a) Total work distribution of the CE protocol. The red dashed line indicates the average work, which is $53.6 ~k_BT$. (b) Work distributions for each particle type: The average works for each particle type are also shown in the figure. A signature of this protocol is that the particles moving from a high potential well to a low potential well involve significantly more work compared to other particle types. Details of the protocol are shown in Figure \ref{table:protocol_table}(a) in Appendix \ref{appendix:detail_of_protocols}.}
\label{fig:work_distribution_for_CE}
\end{figure}

Figure \ref{fig:work_distribution_for_CE}(a) shows the work distributions of all particles at $t_3$ for the protocol described in Figure \ref{fig:Controlled_Erasure_protocol}. The total duration is $150~t_c$ (0.237ns). The full detailed protocol can be found in Figure \ref{table:protocol_table}(a) of Appendix \ref{appendix:detail_of_protocols}. Figure \ref{fig:work_distribution_for_CE}(b) shows the work distributions of each particle type. A distinguishing characteristic of CE work cost distributions is two peaks with a large distance between them. As expected, work is dominated by contributions from the $10$ particles falling into the $00$ well. Because only a quarter of the total particles undergo this transition, the average work for the ensemble is approximately one-fourth of the potential difference $\Delta U_{05}$ between the higher and lower wells. Meanwhile, the peak close to zero represents the $00$, $01$, and $11$ particles, all of which remain in their initial local minima throughout the entire logic cycle. Although some work is done to change their potential energies during the protocol, this work is largely returned to the work reservoir upon completion of a full cycle. Consequently, these particles dissipate relatively minor amounts of energy. 

\subsubsection{Work, Fidelity, and Speed Trade-offs}

An obvious approach to reducing work cost is to decrease $\Delta U_{05}$, the potential difference between the higher and lower wells. This action directly reduces the potential energy converted into kinetic energy as the $10$ particles transition down the potential gradient. This potential difference, however, is inherently constrained by the barrier height separating the $01$ and $11$ particles---$\Delta U_{17}$ and $\Delta U_{37}$. Figures \ref{fig:CE_potential_graphs} (c) and (d) illustrate the potential differences in two distinct scenarios at $t_1$. This comes at the cost of increasing the escape rate for the storage pair and degrades computational fidelity. Under typical operating conditions, $\Delta U_{05}$ is approximately four times greater than both $\Delta U_{17}$ and $\Delta U_{37}$ at the bifurcation point. Given this substantial energy barrier, the probability of the $00$ and $10$ particles thermally exciting into the $10$ region is strongly suppressed. Since they dominate the error, the overall CE fidelity can be accurately characterized by the escape rate of the storage pair, which is exponentially damped in the height of $\Delta U_{17}$ and $\Delta U_{37}$ barriers. Detailed calculations for the escape rate are provided in Appendix \ref{appendix:escape rate}.

This barrier height can be approximated as $\Delta E_B \equiv \beta_1 \cos\frac{\pdc[1]}{2}(1 - \cos\frac{\p[1]^{c}}{2})$, where $\p[1]^{c}$ is the $\p[1]$-coordinate of the $01$ and $11$ potential minima. (See derivation in Appendix \ref{appendix:escape rate}.) In many QFP studies \cite{PhysRevB.46.6338, PhysRevB.80.052506, variable_beta, PhysRevLett.75.1614, Takeuchi_2013, takeuchi2022adiabatic}, it is standard practice to apply the simplification $\pdc[i] \approx \pdc[i]^c \approx \pxdc[i]$, where $\pdc[i]^c$ is the coordinate of the local minima along the $\pdc[i]$ direction. This choice stems from the fact that the parameter $\gamma$ is sufficiently large so that oscillations about the energy minima in $\pdc$ are several times smaller and faster than oscillations in $\p$. Under this assumption, $\Delta E_B$ is calculated by substituting $\pdc[i] = \pxdc[i]$.

\begin{figure}
\includegraphics[scale=0.65]{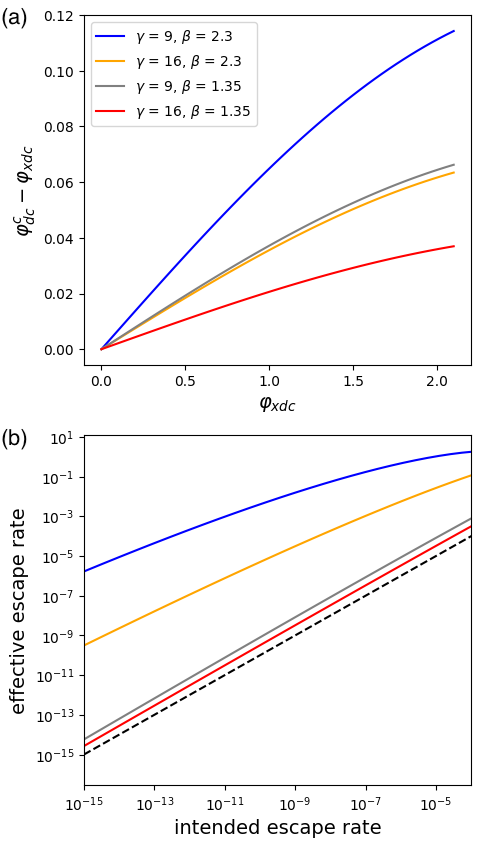}
\caption{(a) This plot shows the difference between $\pdc^c$ and $\pxdc$ as $\pxdc$ increases from 0 to 2.1. The curves are generated by numerically solving Eq. (\ref{equation:phi_dc_vs_phi_xdc}) for four combinations of the coupling parameters, $\beta = \{1.35, 2.3\}$ and $\gamma = \{9, 16\}$. (b) This graph plots the intended escape rate against the effective escape rate per operation. The black dashed line represents the ideal case where the intended escape rate equals the effective escape rate. Note that the effective escape rate is always larger than the intended rate. }
\label{fig:intented_vs_effective}
\end{figure}

However, the exact relation between $\pxdc[i]$ and the local minima $\pdc[i]^c$ can be obtained by taking the first derivative of U with respect to $\pdc[i]$. This yields
\begin{equation}
\pxdc[i] = \pdc[i]^c - \frac{\beta_{i}}{2\gamma_i} \sin\frac{\pdc[i]^c}{2}
  ~.
\label{equation:phi_dc_vs_phi_xdc}
\end{equation}
The true minima $\pdc[i]^c$ is somewhat displaced from $\pxdc[i]$. The numerical solution of this relationship is shown in Figure \ref{fig:intented_vs_effective}(a), which plots the difference $\pdc[i]^c - \pxdc[i]$, as a function of $\pxdc[i]$ for various combinations of $\beta$ = \{1.35, 2.3\} and $\gamma$ = \{9, 16\}. From the plot, the difference is relatively small (no more than 6\%), which appears to justify the assumption $\pxdc[i] \approx \pdc[i]$.

However, this small difference in the coordinate can result in a large difference in the barrier height. We introduce two definitions to distinguish the two barrier-height calculations. \emph{Intended barrier height} refers to that calculated under the assumption $\pdc[i]^c \approx \pxdc[i]$. In contrast, \emph{effective barrier height} is the actual barrier height, which is derived by first solving Eq.(\ref{equation:phi_dc_vs_phi_xdc}) to obtain $\pdc[i]^c$.

We extend these definitions to error rates. The error rate derived from the intended barrier height is defined as the \emph{intended escape rate} and, similarly, the error rate derived from the effective barrier height is defined as the \emph{effective escape rate}. Figure \ref{fig:intented_vs_effective}(b) illustrates the relationship between these two escape rates across different combinations of $\beta$ and $\gamma$. The dashed line in the plot represents the ideal case where the intended and effective escape rates are equal.

Notably, for the case of $\beta = 2.3$, the effective escape rate is observed to be as much as $9$ orders of magnitude higher than the intended escape rate. In contrast, the difference is significantly smaller when $\beta = 1.35$. This startling disparity highlights the fact that even minor variations between $\pxdc$ and $\pdc^c$ can substantially alter the system's energy landscape. This, in turn, leads to drastic implications for the storage pair's escape probability.

\begin{figure}
\includegraphics[scale=0.6]{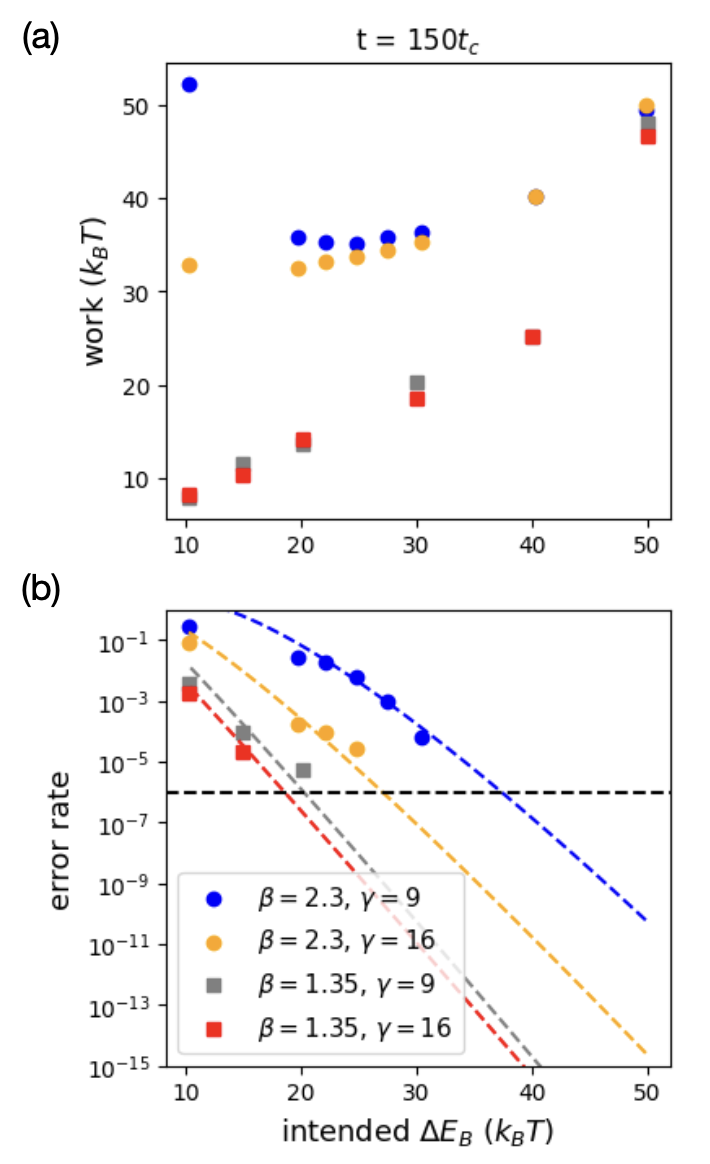}
\caption{(a) Work cost for different combinations of $\beta$ and $\gamma$ with the time for duration with length of 150 $t_c$. The number of samples in each case is $10^6$. When the intended barrier height decreases, the work cost for $\beta = 2.3$ first decreases but then increases afterward. However, for $\beta = 1.35$, the work costs keep decreasing. (b) Error rate against intended barrier height for different combinations of $\beta$ and $\gamma$. The horizontal dashed line represents the threshold of observing errors given the sample size. The colored dashed lines represent the theoretical escape rates for the four cases. The observed error saturates at $.25$, as the particles become effectively randomized in the presence of vanishing barrier height}
\label{fig:work_error_for_CE_t_150}
\end{figure}

Figure \ref{fig:work_error_for_CE_t_150} shows the work cost and fidelity as a function of the intended barrier height for different combinations of $\beta$ and $\gamma$.  One of the most striking features of the simulation results are the substantial differences in the observed error rates for different values of $\beta$ and $\gamma$, even when the ``intended" barrier heights are the same.

This can be explained by the difference between the intended barrier height and the effective barrier height, as described above. The nonmonotonic work cost for $\beta=2.3$ reflects a failure of reliable information storage, also due to the inflated error rate. On the one hand, as the number of errors increases, more and more particles transit between the potential wells and thus do not stay at the potential minima. This increases the work cost. On the other hand, in the case of $\beta = 1.35$ (represented by the gray and red data points), the difference between the intended barrier height and effective barrier height is much smaller. Although errors are also observed for intended barrier height smaller than $20 ~k_BT$, the error rate is orders of magnitude smaller.

\begin{figure}
\includegraphics[scale=0.7]{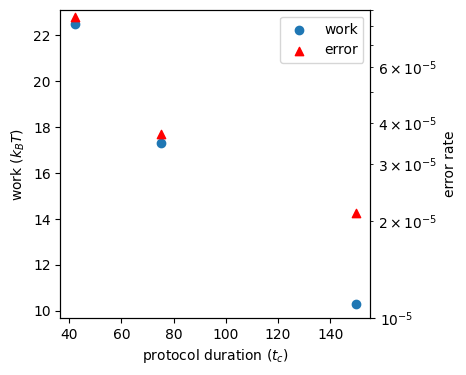}
\caption{Work cost and error rate for a system with $\beta = 1.35$ and $\gamma = 16$ and with intended barrier height $15~k_BT$. Simulated for three different protocol durations: $42$, $75$, and $150$ $t_c$. As speed increases, the work cost and error rate increase.}
\label{fig:work_speed_and_error}
\end{figure}

Our analysis now shifts to the effect of operating speed on work cost and fidelity. Figure \ref{fig:work_speed_and_error} illustrates the relationship between total protocol duration and performance, using a base case with $\beta=1.35$ and $\gamma = 16$ and an intended barrier height of around $15~k_BT$. Both work cost and the error rate increase as the total protocol duration is shortened from $150~t_c$ to $42~t_c$. A faster protocol gives particles extra kinetic energy through sudden changes in potential energy. This excess energy increases the probability of the storage pair escaping its intended wells, leading to a higher error rate. Additionally, it increases the work cost because the excess kinetic energy is eventually dissipated into the environment.  To maintain fidelity and reduce escape events as the barrier height decreases, the protocol must be executed slowly enough to ensure that the storage particles remain locally equilibrated to their local minima throughout the operation.


\subsubsection{Problem of Controlled Erasure}

In the CE protocol, maintaining a high potential barrier between the $01$ and $11$ wells is essential to prevent undesired particle transitions. First, this requirement---maintaining the storage pair within their own wells---led us to employ a nonadiabatic saddle-node bifurcation, rather than an adiabatic pitchfork bifurcation. Next, we observed another direct trade-off stemming from the storage bits: reducing the barrier height lowers the work cost but simultaneously decreases fidelity by facilitating particle escape events. Another critical limitation we identified is that the operating speed of a high-fidelity CE is constrained by the storage pair. As the protocol's speed increased, these particles gain kinetic energy, causing them to escape their storage wells and compromise the outcome.

This highlights a fundamental constraint on CE: It relies on static particle positioning to store information during the computation. Theoretically, this should be a strength, since particles near their local equilibrium distributions dissipate energy very slowly. However, since the relevant barrier heights in this system are inescapably coupled during a CE, the result is a tension between the storage pair and the erasure pair. In short, the storage pair needs a constant passive barrier and the erasure pair needs a vanishing one.

The solution is to explore a method that does not require particles of the storage pair to be static. The next section introduces a new protocol, the Erasure-Flip protocol (EF), to solve this problems.

\subsection{Mechanism of Erasure-Flip Protocol}
\label{section:erasure_flip_protocol}

In the momentum computing bit-flip protocol \cite{ray2022gigahertz}, the potential is abruptly changed from a double-well to a quasi-harmonic potential. This harmonic potential is maintained for half an oscillation period to allow the particles representing logical $0$ and logical $1$ to flip their positions. The potential then reverts to its original double-well form, trapping the flipped particles and preventing them from returning to their initial wells. A key feature of this process is that particles remain separable even when they are spatially mixed near the bottom of the harmonic potential, as their distinct start states are preserved in their opposing momenta. This ability to store and manipulate information in momentum space is a powerful feature not possible with traditional overdamped Langevin dynamics.

Inspired by this, the EF also utilizes a momentum-computing approach. Figure \ref{fig:truth_table_and_three_operations} displays the truth table for the EF protocol, along with the initial and final particle distributions for a full EF cycle. In contrast to the CE, the $00$ and $10$ particles merge within the $10$ well, while the $01$ and $11$ particles flip their positions. Despite these differences, we maintain the nomenclature of calling the $00$ and $10$ particles the erasure pair and the $01$ and $11$ particles the storage pair. After all, the flip operation is logically reversible. And so, the logical information separating $01$ and $11$ particles is preserved.

\begin{figure*}
\includegraphics[scale=0.44]{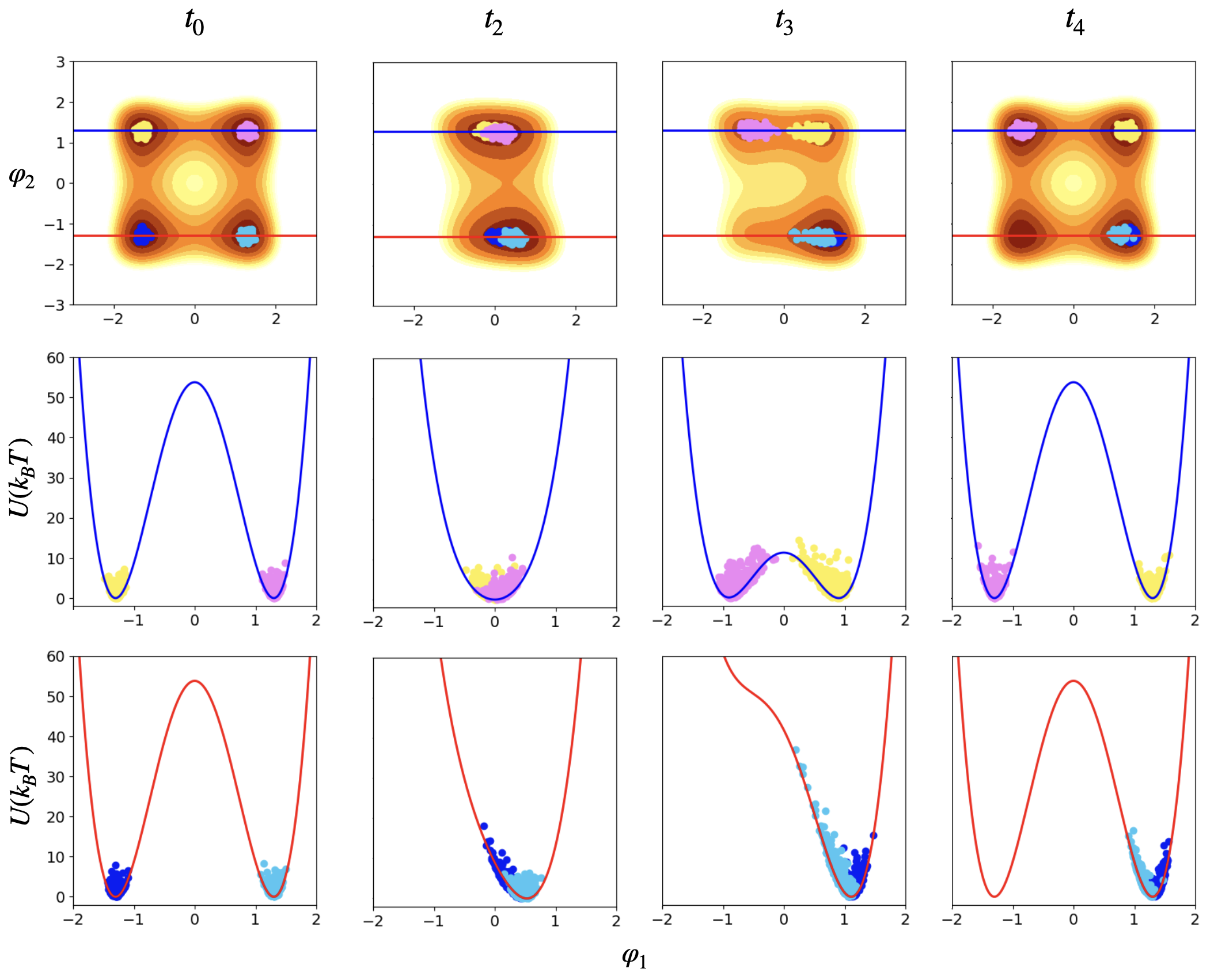}
\caption{Snapshots of the evolution of the potential and particle distributions of the EF protocol at four important time steps: $t_0,t_2,t_3, \text{and } t_4$. The partial barrier lowering at $t_1$ is not shown. (\textcolor{blue}{\href{https://drive.google.com/file/d/1t8FjEA-MYU0axwYPkKIDzbjpKlF0Mtmh/view?usp=sharing}{Animation available online}}).}
\label{fig:Erasure_Flip_protocol}
\end{figure*}


Figure \ref{fig:Erasure_Flip_protocol} shows the distribution of the particles and the potential energy of the system at four different times during an EF protocol. The protocol divides into three main stages. On the one hand, in the first stage ($t_0$ to $t_2$), the potential along the erasure pair is tilted toward $10$ well so that $00$ moves toward the $10$ well. On the other hand, the barrier height between the storage pair is reduced to zero. The resulting quasi-harmonic potential allows the $01$ and $11$ particles to flip with each other. In the second stage ($t_2$ to $t_3$), the potential barrier is raised to trap these flipped particles, preventing them returning to the previous well. In the last stage ($t_3$ to $t_4$), the potential returns to the original four-well potential. 

In this way, the EF's design addresses the CE's fundamental limitations. In stark contrast to CE, which must avoid the storage pair needlessly gaining kinetic energy, we intentionally impart kinetic energy to them and encourage these particles to move into the other well. Thus, the excess kinetic energy that was a source of error in CE is now a necessary condition for successful EF.

\subsection{Simulation Results for EF}

\subsubsection{Work distributions}

\begin{figure*}
\includegraphics[scale=0.57]{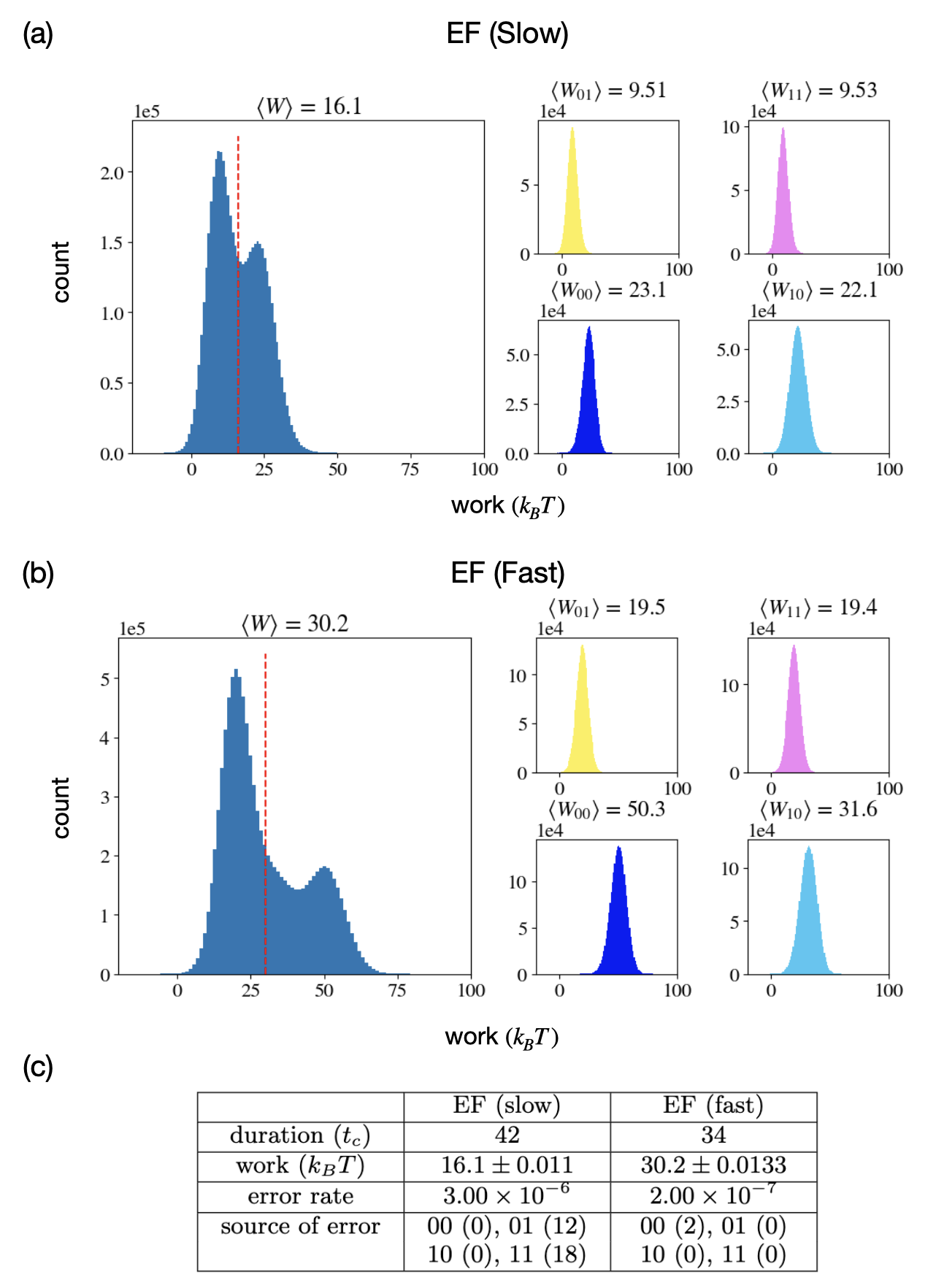}
\caption{Work distributions for (a) EF (slow) and (b) EF (fast). For each protocol, the total work distribution across all particles and the work contribution from individual particle types are shown. The total number of particles simulated for both protocols is $N=10^7$. (c) A summary table illustrating the duration, mean work cost, error rate, and source of error for the two protocols. For the error source, the number in the bracket is the number of observed errors for that particular particle type. Details of the protocol parameters are shown in Figure \ref{table:protocol_table}(b) and (c).}
\label{fig:work_distribution_for_EF_1}
\end{figure*}

Figure \ref{fig:work_distribution_for_EF_1}(a) shows the total work for the EF protocol and individual work distributions for each particle type. In contrast to the CE, all four types move under EF. As none of them simply sit near local minima during the protocol, we expect for each particle type that some input work will change the potential energy and so be converted to kinetic energy and dissipated to the surrounding. The average work cost is about $16~k_BT$, with the average cost of the storage and erasure pairs being $\approx 9~k_BT \text{ and }23~k_BT$, respectively. Here, the storage pair costs relatively less work since a significant portion of the kinetic energy gained from $t_0$ to $t_2$ is reinvested into potential energy from $t_2$ to $t_3$.

Section \ref{section: controlled_eraure}'s CE analysis demonstrates that lowering the potential barrier height is beneficial for reducing the potential difference between the higher and lower wells of the erasure pair, thereby decreasing work cost. On the one hand, the EF takes this principle further by completely removing the potential barrier, allowing for a lowering the amount of potential energy converted to kinetic energy when $00$ particles slide down the CE potential gradient. On the other hand, the EF protocol requires a rapid change in $\pxdc$ to trap the particles at just the right time. The work cost associated with this rapid change is directly related to the parameter $\gamma$, which can be conceptualized as the system's ``stiffness'' in the $\pdc$ space. Consequently, a smaller value of $\gamma$ is crucial for reducing the work cost of EF.

\begin{figure*} 
\includegraphics[scale=0.4]{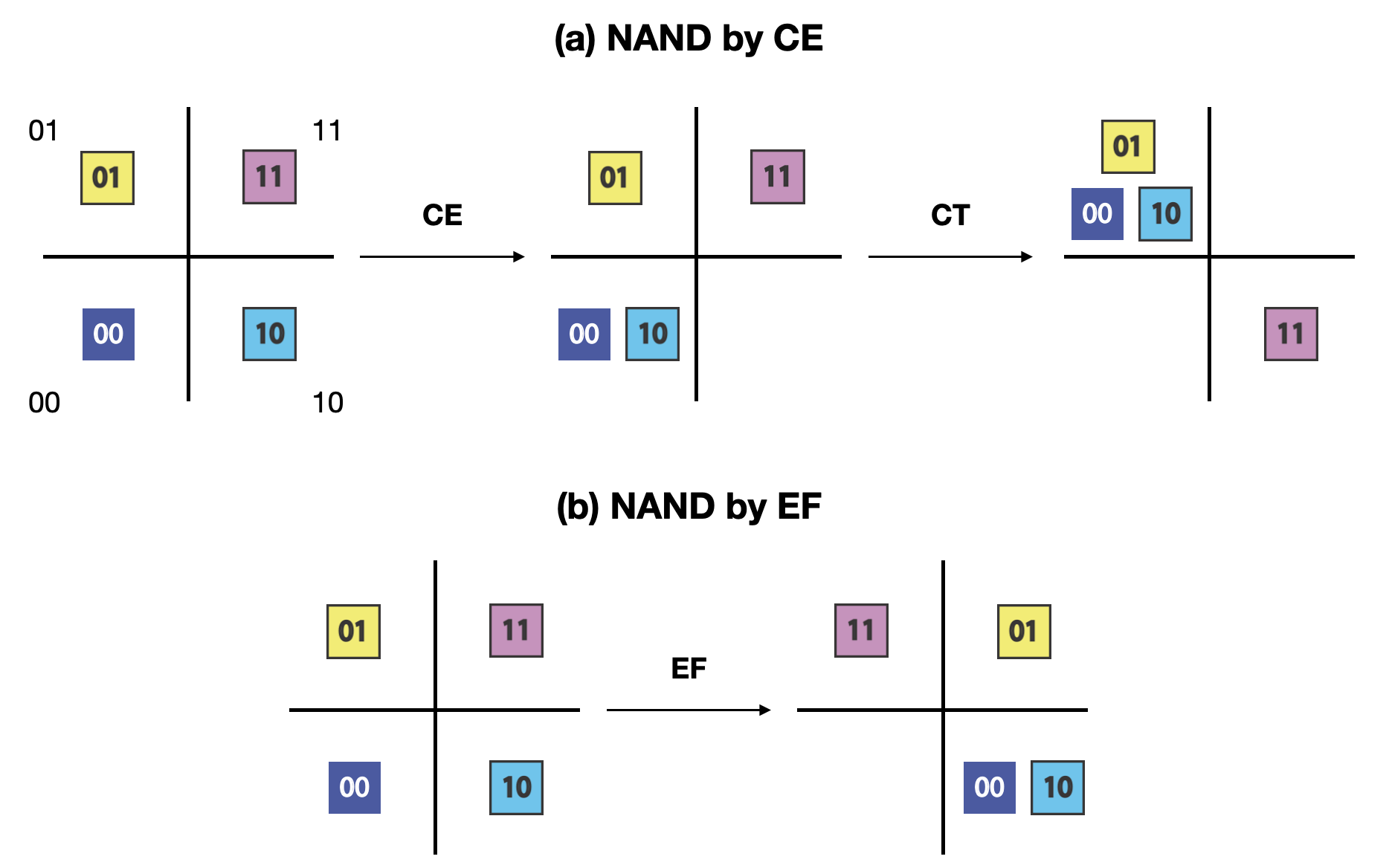}
\caption{Schematic illustration comparing the partial NAND operation under (a) CE and (b) EF. After a CE operation, one more conditional tilt operation (CT) is required to finish the NAND operation. The first bit is the irrelevant and the second bit is the output bit. For the EF, the final distribution already satisfies the requirement of a partial NAND. The first bit is the output bit and the second bit is the irrelevant bit.}
\label{fig:NAND_by_CE_and_EF}
\end{figure*}  

\subsubsection{Trading-off work, fidelity, and speed}

With the potential barrier from the $10$ well to the $00$ well exceeding $50~k_BT$ at $t_3$ (Fig. \ref{fig:Erasure_Flip_protocol}), the probability of a particle transitioning back to the $00$ well becomes exceedingly low. Consequently, the primary source of error still comes from the storage pair---specifically from the flipping  between the $01$ and $11$ particles. The mechanism behind these errors is, however, quite different from CE.

The precise timing of the barrier-lowering phase is critical to the EF protocol's success. If the barrier is lowered too slowly ($t_0$ to $t_2$), particles may not gain sufficient kinetic energy to transition to the target well. Instead, they could remain near the central point ($\p[1] = 0$) when the barrier starts to rise and end up in a random (indeed, wrong) well. Conversely, if the barrier is raised too quickly ($t_2$ to $t_3$), some particles may not have reached the target well. Then rising the barrier too slowly could allow particles in the target well to return to their original well. Therefore, a careful balance (and protocol timing) must be maintained to optimize both work cost and fidelity.

The simulation error rate and error source are summarized in the table of Figure \ref{fig:work_distribution_for_EF_1}. In $10^7$ simulations, we observed only $30$ instances of particles failing to reach the target wells, yielding an error rate of $3 \times 10^{-6}$. As expected, all of the failure particles are associated with the storage pair ($12$ trajectories from $01$ particles and $18$ trajectories from $11$ particles) because these particles do not have sufficient speed to complete the transition.

Therefore, we simulated an alternative, faster protocol. Figure \ref{fig:work_distribution_for_EF_1}(b) shows the results. Across $10^7$ trajectories, only two errors were observed, and these originated from the erasure pair rather than the storage pair. A detailed comparison of the average kinetic energy and examples of failure trajectories for the slow and fast EF are shown in Appendix \ref{appendix:slow_EF_vs_fast_EF}. This comparison suggests an intermediate protocol lying between these two might best minimize the error rate.

Taken together, these results highlight that a delicate balance must be struck between protocol parameters and duration to find optimal momentum-computing protocols. Finding an optimal protocol for EF that balances speed, work cost, and fidelity remains an open question. To address this, though, we have developed machine learning methods \cite{lyu2025} that are a promising and well-behaved approach for optimization.

\subsection{NAND operation by Controlled Erasure and Erasure-Flip}
\label{section:NAND_protocol}

Finally, we discuss the costs of universal computation using both the adiabatic and the momentum computing approaches. Figure \ref{fig:NAND_by_CE_and_EF} shows the steps for achieving a partial NAND operation using either the CE or the EF protocol. A NAND based on the CE requires an additional operation. In order to stay within the adiabatic paradigm, we choose an operation that does not take advantage of momentum: the conditional tilt (CT). A CT is done first by lowering the potential barriers along a particular direction, and then changing the magnetic coupling. This protocol was investigated in detail in \cite{Tang_2025}; it can be implemented adiabatically---with work costs that approach Landauer's bound. While the energy cost of a CT can theoretically approach the Landauer limit with an infinitely long duration, we operate at a finite time scale, so the additional costs levied by this step are an important factor. Note also that, in this case, the first output bit is the irrelevant bit and the second output bit yields the NAND operation.


\begin{figure*} 
\includegraphics[scale=0.55]{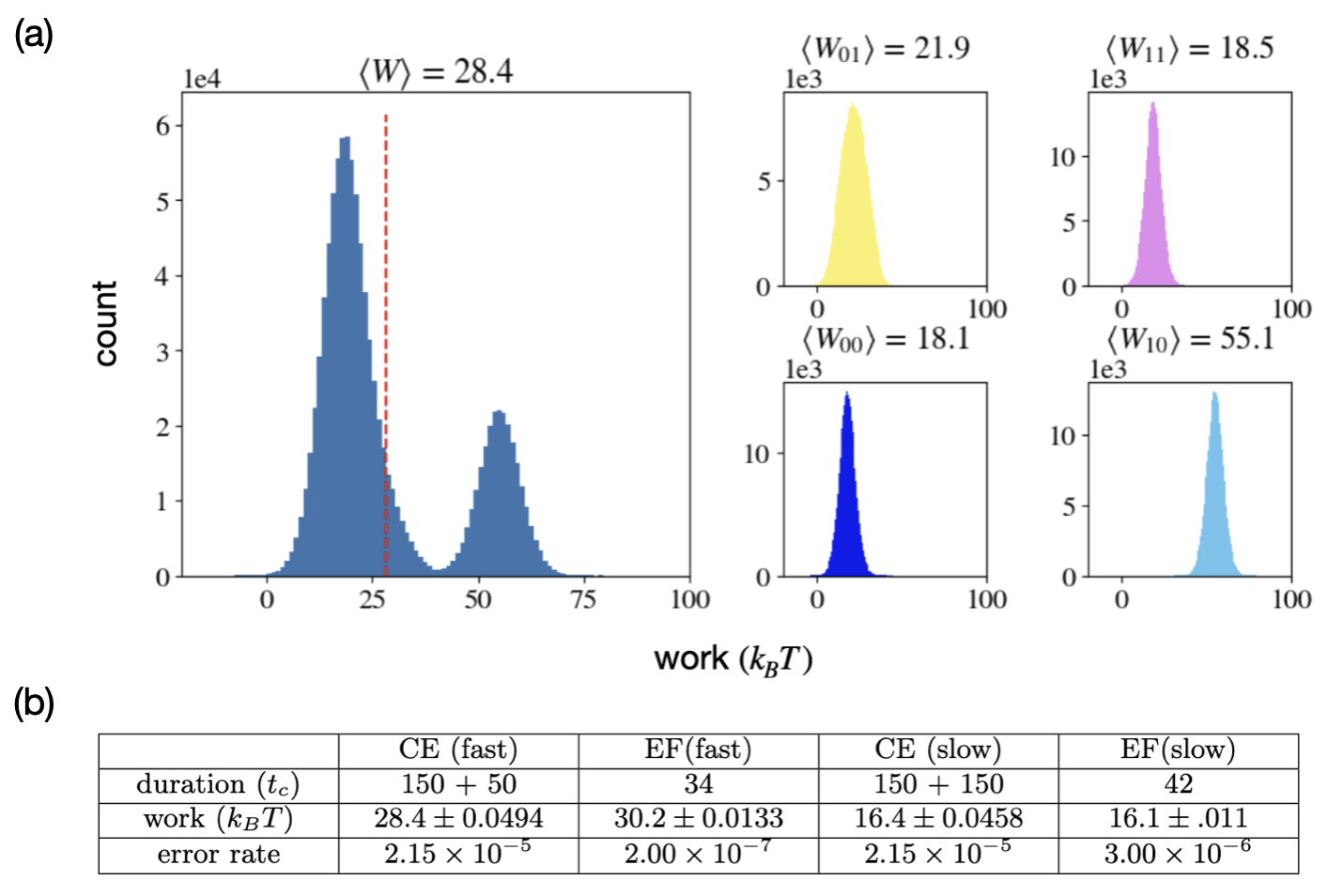}
\caption{(a) This panel shows the total work distribution alongside the individual particle-type work distributions for the partial NAND protocol implemented via CE with CT duration $50~t_c$ (CE (fast)). Since the shape of the work distribution of CE (slow) is the same as CE (fast) with lower work cost, the work distribution for CE (slow) is not shown here. Animation for the protocol available via the links: \textcolor{blue}{\href{https://drive.google.com/file/d/1Cg9b72_wYZKOxwxagG6SkjLKFxtTtRk7/view?usp=sharing}{NAND by CE (fast)}}. (b) Table listing key performance metrics, including the duration, mean work cost, and error rate for CE (fast), EF (fast), CE (slow) and EF (slow) schemes.}
\label{fig:work_distribution_of_NAND_by_CE_and_EF}
\end{figure*}  

Figure \ref{fig:work_cost_of_CT} in Appendix \ref{appendix:detail_of_protocols} details the work cost of CT as a function of its duration. We find that a CT with a duration of $50t_c$ incurs an energy cost of approximately $20 k_BT$. The cost reduces to around $4k_BT$ when the duration is extended to $250t_c$. To facilitate a meaningful thermodynamic comparison, we select CT durations to ensure that the work cost of the NAND operations implemented by CE are comparable to the work costs of their respective counterparts, the EF (fast) and EF (slow) protocols. This results in CT durations of $50t_c$ and 150$t_c$; we designate the respective CE protocols as CE (fast) and CE (slow), respectively.
 
 On the other hand, a single EF cycle is sufficient to implement a partial NAND operation; the first output bit can be read directly to get the NAND result. The second output bit is simply a copy of the second input bit.

Figure \ref{fig:work_distribution_of_NAND_by_CE_and_EF} illustrates the work cost achieved when implementing the NAND operation using the CE (fast) protocol. The work distributions for the NAND operation implemented with the EF protocols are already presented in Figure \ref{fig:work_distribution_for_EF_1}, because no extra steps are required. The accompanying table summarizes the key comparative metrics---work cost, fidelity, and speed---for the four main protocols. Considering the comparison between NAND by CE (fast) and NAND by EF (fast), the advantage of EF protocol becomes apparent. While the mean work costs for both protocols are approximately $30~k_BT$, EF (fast) protocol demonstrates vastly superior performance in both speed and fidelity. Specifically, EF (fast) implementation is six times faster and achieves an improvement in fidelity of two orders of magnitude compared to CE. Furthermore, comparing CE (slow) with EF (slow) also demonstrates EF's better overall performance. With work cost roughly $16~k_BT$, EF (slow) is approximately seven times faster and the fidelity is ten times better than that of CE (slow). 

As a final point of comparison, we speed up the CE-based NAND so that it is as fast as the fastest EF with a duration of $34~t_c$. This induces a huge energy penalty with an average work of $\approx108~k_B T$ and the error rate spikes to an untenable $\approx 13\%$ (for comparison, randomization of unbiased inputs would achieve a partial NAND $50\%$ of the time). At comparable operating speeds, it costs $3.6$ times more energy while incurring more than $650,000$ errors for every error seen in the EF-based implementation. These comparisons underscore the foundational advantage of our hybrid reversible/irreversible logic as a vehicle for general-purpose computing. The remarkable combination of speed and accuracy, achieved across different operating points without increasing thermodynamic costs, demonstrates the profound advantage of leveraging momentum in computation. And, this points to promising pathways for achieving dramatic improvements in speed, energy efficiency, and fidelity in digital computation.

\section{Discussion}
\label{section:conclusion}

This investigaton first pursued a detailed analysis of the work, fidelity, and speed of the Controlled Erasure (CE) protocol. The findings revealed an inherent lower bound on its work cost, when implemented with a CQFP. This cost arises from the unavoidable transfer of particles from higher to lower potential wells, due to an inescapable use of a saddle-node bifurcation. Furthermore, the protocol's speed was fundamentally limited by the storage pair, as rapid execution generates excessive kinetic energy, which in turn increases the probability of escape events.

To overcome these limitations, we introduced a logic gate that simultaneously performs reversible and irreversible logics on different memory subspaces. This universal operation, the Erasure-Flip (EF), transiently stores information in momentum space during the computational protocol, allowing logical states to depart from their local equilibria without losing information. This flexibility overcame the speed limit imposed by requiring information be stored in near-equilibrium quasi-static states. While the EF protocol does present greater control challenges, our results show the EF protocol significantly outperforms our adiabatic test operation in operating speed and in computational fidelity without leveraging additional energetic penalties. This successfully demonstrates the strength of the underdamped momentum computing paradigm when compared to the conventional overdamped approach. This finding opens the door to fundamentally different and more efficient computational architectures. This, in turn, paves the way for highly energy-efficient, accurate, and fast protocols that are not achievable with traditional overdamped perspective. Future work will focus on further exploring and building upon this promising new thermodynamic computing paradigm.

\section{Method}

The simulation is written in Python using its NumPy library, and the figures and animations are generated by Matplotlib. To obtain the initial states, Monte Carlo sampling was performed. The system dynamics were then simulated using a hybrid integration scheme: a 4th-order Runge-Kutta method was applied to the deterministic portion, while the Euler-Maruyama method was used for the stochastic portion. This procedure generates an ensemble of trajectories for the system's initial state. Physical quantities, such as the mean work, are computed by taking the average across the trajectories produced by the simulator. A fixed time step of $dt = 1/1000$ was employed for all simulations. Statistics for physical observables, such as the mean work, are obtained by ensemble averaging over independent trajectories. Detailed derivations of the Langevin equation and additional simulation parameters are provided in Appendices \ref{appendix:langevin_dynamics} and \ref{appendix:detail_of_the_simulation}.

\section{Data Availability}

The data in the manuscript are provided in the main text, Supplementary Note, and available code repository. Additional data related to this study can be made available from the authors upon reasonable request.

\section{Code availability}

The simulator used is open-sourced and available at https://github.com/tkwtang/source/.

\section{Author Contributions}

KWT conceived the idea of the Erasure-Flip and was responsible for generating the simulation data, formal analysis, and visualization. KJR and JPC provided critical feedback, verified the theoretical framework, and were involved in supervision. JPC provided funding. All authors contributed to the editing of the manuscript and approved the final version for submission.

\section{Competing Interests}

The authors declare no competing interests.

\begin{acknowledgments}
We thank Christian Pratt and Sangbum Kim for helpful discussions. The authors thank the Telluride Science Research Center for its hospitality during visits and the participants of the Information Engines workshop there for their valuable feedback. This material is based on work supported by, or in part by, the Art and Science Laboratory via a gift to UC Davis.
\end{acknowledgments}

\bibliography{references.bib}
\clearpage
\appendix

\section{Langevin Dynamics for Josephson Circuits}
\label{appendix:langevin_dynamics}

Underdamped Langevin dynamics can often be used to simulate the behavior of the flux degrees of freedom when Josephson junctions (JJs) are modeled as 
resistively and capacitively shunted junction (RCSJ) model. The dimensional Langevin equation is:
\begin{equation}
\label{langevin_equation_1}
m_i dv_i + \nu_i v_i dt = -\partial_{x_i}
U(\vec{x}; \vec{\lambda}_{ext}(t)) dt + r_i(t) \sqrt{2 \nu_i \kappa dt}
  ~.
\end{equation}
Here, $U(\vec{x}; \vec{\lambda}_{ext}(t))$ is the driving potential and $\vec{\lambda}_{ext}(t)$ is the protocol of the external parameters. $\kappa$ is equal to $k_BT$. The symbols $x_i$, $v_i$, $m_i$, $\nu_i$, and $r_i$ represent the $i$-th components of the $\vec{x}$ (position), $\vec{v}$ (velocity), $\vec{m}$ (mass), $\vec{\nu}$ (damping), and $\vec{r}$ (Gaussian random number), respectively In this, the $r_i(t)$ are the 4 independent memoryless Gaussian random variables with zero mean and unit variance. Note that $r_i(t)$ have the following correlation:
\begin{align}\label{etaTimeCorrelations}
\langle r_i(t) r_j(t') \rangle = \delta_{ij}\delta(t-t').
\end{align}
The average $\langle \ldots \rangle$ is taken over the stochastic process trajectories.

The following shows how to obtain the dimensionless Langevin equations from the above dimensional equation. Rearranging terms of Eq. (\ref{langevin_equation_1}) gives:
\begin{align} 
\label{eq:langevin_equation_2}
dv_i = -\frac{\nu_i}{m_i} v_i dt - \frac{1}{m_i} \partial_{x_i} U(\vec{x})dt + \frac{r_i(t)}{m_i} \sqrt{2 \nu_i \kappa dt}
\end{align} 

These quantities can be written as a product of a constant with unit and a unitless quantity: $\vec{x} = x_c \vec{x'}, \vec{v} = v_c \vec{v'}, \vec{m} = m_c \vec{m'}, \vec{\nu_i} = \nu_c \vec{\nu_i'}, U = U_0 U', \kappa = \kappa_c \kappa'$ and $t = t_c t'$. Substitute these expression into Eq. (\ref{eq:langevin_equation_2}), it can be expressed as:
\begin{multline}
    v_c dv' = - \frac{\nu_c \nu'}{m_c m'} v_c v' t_c dt' - \frac{t_c}{m_cm'} \frac{U_0 }{x_c} \partial_{x'} U' dt' \\ 
    + \frac{1}{m_c m'}r(t') \sqrt{2 \nu_c \nu' \kappa_c \kappa' t_c dt'}.
\end{multline}
Note that subscript $i$ is dropped to simplify the expression.
Dividing by $v_c$ on both sides and making use of $v_c = x_c/t_c$, the above can be simplified:
\begin{multline}
    dv' = - \frac{\nu_c t_c}{m_c } \frac{v'}{m'} v' dt' - \frac{1}{m'}  \partial_{x'} U' dt' \\
    + \frac{1}{m_c}  \sqrt{ \frac{\nu_c \kappa_c t_c^3} {x_c^2} } \frac{\sqrt{\nu' \kappa'}}{m'} r(t') \sqrt{2dt'}
    ~.
\end{multline}
This can then be further simplified to: 
\begin{align} \label{langevin_equation_3}
dv' = - \lambda v' dt' - \theta \partial_{x'} U'dt' + \eta r(t') \sqrt{2dt'}
    ~.
\end{align}
Here, $\lambda$ is the thermal coupling coefficient, reflecting the strength of the damping force the particles experience from the thermal bath. $\theta$ is related to the strength of potential relative to the dynamic of the system. If $\theta$ is zero, the system is off from the potential. $\eta$ determines noise strength. Returning subscript $i$ gives the dynamics in the dimensionless form of the Langevin equations:
\begin{equation} \label{electrostatic}
dv'_i = - \lambda_i v'_i dt' - \theta_i \partial_{x'_i} U' dt' + \eta_i r_i(t') \sqrt{2dt'}
  ~,
\end{equation}
where $\lambda_i, \theta_i$, and $\eta_i$ are defined as:
\begin{align*}
\lambda_i &=  \frac{\nu_c t_c}{m_c } \frac{\nu_i'}{m_i'} ,\\
\theta_i &= \frac{1}{m_i'}, ~\text{and}\\
\eta_i &= \frac{1}{m_c}  \sqrt{ \frac{\nu_c \kappa_c t_c^3} {x_c^2} } \frac{\sqrt{\nu' \kappa'}}{m'} = \sqrt{\frac{\lambda_i \kappa'}{m'}}
  ~.
\end{align*}

In our specific circuit, with the assumption that $L = L_1 = L_2$, $C = C_1 = C_2$ and $R = R_1 = R_2$, the vectors can be written as:
\begin{align*} 
\vec{x} &= (\phi_1, \phi_2, \phi_{1dc}, \phi_{2dc}) ~,\\
\vec{v} &= (\frac{d \phi_1}{dt}, \frac{d \phi_2}{dt}, \frac{d \phi_{1dc}}{dt}, \frac{d \phi_{2dc}}{dt})~,\\
\vec{m} &= (C, C, \frac{C}{4}, \frac{C}{4})~,\\
\vec{\nu_i} &= (\frac{2}{R}, \frac{2}{R}, \frac{1}{2R}, \frac{1}{2R})~, ~\text{and}\\
\vec{r} &= (r_1, r_2, r_3, r_4)
  ~.
\end{align*}
Choosing the following constants,
\begin{align*} 
x_c &= \frac{\Phi_0}{2\pi}; v_c = \frac{x_c}{t_c}; t_c = \sqrt{LC};\\
m_c &=C; \nu_c = \frac{1}{R}; ~\text{and}\\
\kappa_c &= U_0 = \frac{m_c x_c^2}{t_c^2} = \frac{\Phi_0^2}{4 \pi^2 L},
\end{align*}
the vectors then become:
\begin{align*} 
\vec{x} &= x_c \vec{x'} = \frac{\Phi_0}{2\pi} (\p[1], \p[2], \pdc[1], \pdc[2]),\\
\vec{v} &= v_c \vec{v'} = \frac{x_c}{t_c} (\frac{d\p[1]}{dt}, \frac{d\p[2]}{dt}, \frac{d\pdc[1]}{dt}, \frac{d\pdc[2]}{dt}),\\
\vec{m} &= m_c \vec{m'} = C (1, 1, \frac{1}{4}, \frac{1}{4}), ~\text{and}\\
\vec{\nu_i} &= \nu_c \vec{\nu_i'} = \frac{1}{R} (2, 2, \frac{1}{2}, \frac{1}{2}),
  ~
\end{align*}
Therefore, $\vec{\lambda}, \vec{\theta}$ and $\vec{\eta}$ can be expressed:
\begin{align*}
\vec{\lambda} &= \frac{2}{R} \sqrt{\frac{L}{C}}(1, 1, 1, 1) ,\\
\vec{\theta}  &= (1, 1, 4, 4), ~\text{and}\\
\vec{\eta}    &= \sqrt{\frac{k_BT }{U_0}} (\sqrt{\lambda_1}, \sqrt{\lambda_2}, 2\sqrt{\lambda_3}, 2\sqrt{\lambda_4})
  ~.
\end{align*}

\section{Escape rate calculation}
\label{appendix:escape rate}

\subsection{Escape rate}

The escape rate controls the number of escape events in one second per particle. The escape rate for 2D \cite{PhysRevLett.63.1712, PhysRevB.46.6338, PhysRevB.35.4682} is related to the barrier height $\Delta E_B$ by the following equation:
\begin{align}
    \Gamma = a_t \frac{\Omega}{2\pi}\exp \left(-\frac{\Delta E_B}{k_BT} \right).
\end{align}
Here, $\Omega = \frac{\omega_{lw}\omega_{tw}}{\omega_{ls}}$, where $\omega_{lw}$ and $\omega_{tw}$ are the angular frequencies of the oscillations in the longitudinal and transverse directions at the local minima, and $\omega_{ls}$  is the angular frequency of the oscillations in the longitudinal direction at the saddle point. Here, longitudinal (transverse) direction means the direction parallel (perpendicular) to the oscillation direction and transverse  Using the second derivative approximation for oscillation frequency, these angular frequencies can be calculated using:
\begin{align*}
\omega = \frac{U_0}{x_c^2}\frac{d^2U'}{d\p^2}
\end{align*}
at points of interest. $\Delta E_B$ is the barrier height separating the local minima and saddle point. The theoretical number of escape events per operation is: 
\begin{equation*}
    \Gamma_{cycle} = \Gamma \times N \times t_{tot},
\end{equation*}
where N is the number of particles in the simulation and $t_{tot}$ is the duration of the protocol. 

\subsection{Barrier height between $01$ and $11$ particles}

The barrier height is calculated by finding the difference of the potential at the top of the barrier height, located at ($\p[1], \p[2])= (0, \p[2]^c$), and the potential at the bottom of the well, located at $(\p[1], \p[2])= (\pm \p[1]^{c}, \p[2]^c$), where $\p[1]^{c}$ and $\p[2]^c$ are the absolute values of the critical points in $\p[1]$ and $\p[2]$ respectively. Thus, the barrier height between 01 and 11 particles ($\Delta E_B$) can be approximated as:
\begin{align*}
    \Delta E_B &= U(\p[1] = 0, \p[2] = \p[2]^c)  - U(\p[1]=\p[1]^{c}, \p[2] = \p[2]^c)\\
             &= \beta_1 \cos\frac{\pdc[1]}{2}(1 - \cos\frac{\p[1]^{c}}{2}) + \frac{\xi}{2}[(\px[1])^2- \\
             & (\p[1]^{c} - \px[1])^2 ] - m_{12}\xi\p[1]^{c}(\p[2]^c - \px[2]), 
\end{align*}
where $U$ is the potential $U$ in Eq. (\ref{eq:QFP_potential}). However, for the storage pairs, because the second and third terms balance out, $\Delta E_B$ is reduced to $\beta_1 \cos\frac{\pdc[1]}{2}(1 - \cos\frac{\p[1]^{c}}{2})$.


\section{Simulation Details}
\label{appendix:detail_of_the_simulation}

\subsection{Parameters}

Table \ref{table:circuit_parameters} summarizes the simulation parameters. Once the circuit is fabricated, the fabrication parameters are set. Consequently, the calculated parameters are entirely dependent upon these established fabrication parameters. The parameter $\delta \beta_i$ describes how asymmetric the potential is. This is set to be zero to keep the simulation simple, but in practice will be nonzero based on fabrication variability. Some amount of asymmetry can be compensated through calibrating offsets to the $\px[i]$ control values, though an ideally symmetric device is preferable.

\begin{figure*}
\includegraphics[scale=0.6]{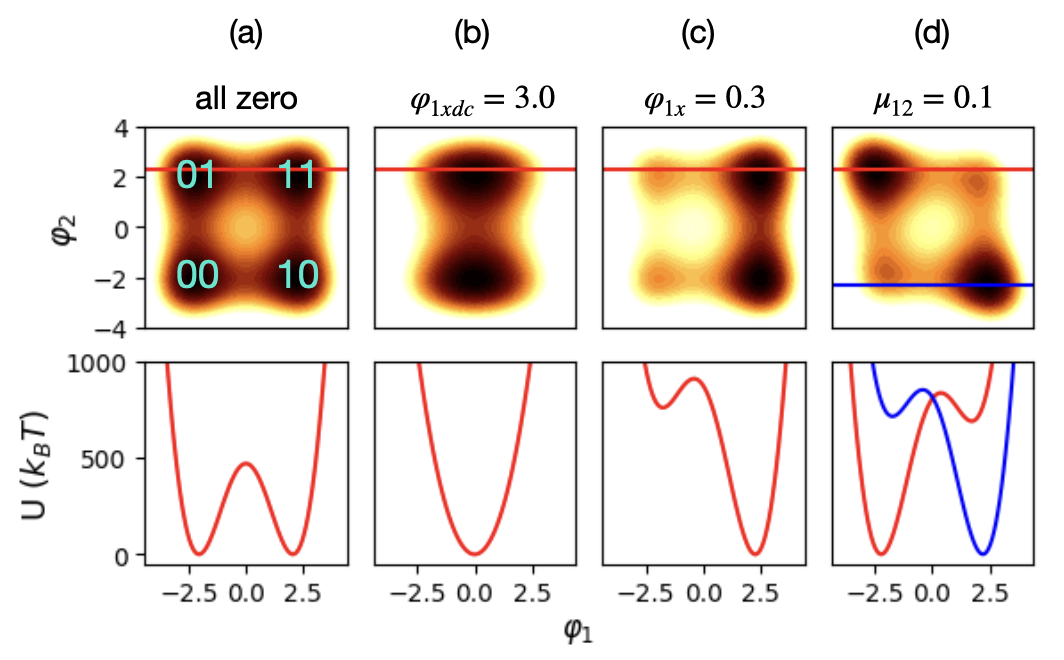}
\caption{Potential landscape in the space of $\p[1]$ and $\p[2]$: The line graphs below show the energy profile along the red/orange line in the contour plots. (a) When all external parameters $\px[1], \px[2], \pxdc[1], \pxdc[2]$, and $m_{12}$ are zero, the energy landscape is a $4$-well potential. The $2$-digit numbers in the contour plot---(00, 01, 10, 11)---label the four wells. (b) $\p[ixdc]$ controls the barrier heights between potential wells. For simplicity, only the case of $\p[1xdc]$ is shown in this figure. (c) $\p[ix]$ tilts the potential. For simplicity, only the case of $\p[1x]$ is shown in this figure. (d) $m_{12}$ conditionally tilts the potential landscape based on particle location in the left or right half-plane.}
\label{fig:parameter_functions}
\end{figure*}

The external parameters are controls for modifying the potential landscape and interacting with the system dynamics. Figure \ref{fig:parameter_functions} shows the effect of the external parameters on the potential landscape. When all external parameters are set to zero, the system exhibits four distinct local minima, forming a $4$-well potential (Figure \ref{fig:parameter_functions}a). $\pxdc[1(2)]$ adjust the potential barriers between top (left) and bottom (right) wells, while $\px[1(2)]$ tilts the potential landscape horizontally (vertically). For simplicity, only the horizontal cases are shown in (Figure \ref{fig:parameter_functions}b and c). $m_{12}$ conditonally tilts the potential landscape (Figure \ref{fig:parameter_functions}d).

\begin{table} 
\begin{subtable}{(a) Fabrication parameters}
\centering
\begin{ruledtabular}
\begin{tabular}{l l l}
\textrm{Symbol}&
\textrm{Physical meaning}&
\textrm{Value}\\
\colrule
 $R_1, R_2$ & Resistance of the JJs & $ 100 \Omega$ \\
 $C_1, C_2$ & Capacitance of the JJs & 1 pF \\
 $L_1, L_2$ & Inductance of the rf-SQUIDs & 5 pH \\
 $l_1, l_2$ & Inductance in series with the JJs & 0.16 - 0.5pH \\
 $I^c_{1a}, I^c_{1b}$ & Critical currents of $J_{1a}$ and $J_{1b}$ & $45, 75 ~\mu A$\\
 $I^c_{2a}, I^c_{2b}$ & Critical currents of $J_{2a}$ and $J_{2b}$ & $45, 75 ~\mu A$\\
\end{tabular}
\end{ruledtabular}
\end{subtable}

\begin{subtable}{(b) Calculated parameters}
\centering
\begin{ruledtabular}
\begin{tabular}{l l l}
\textrm{Symbol}&
\textrm{Formula}&
\textrm{Value}\\
\colrule
 $I_{+1}, I_{+2}$ & $I^c_{1a} + I^c_{1b}$, $I^c_{2a} + I^c_{2b}$ & $90, 150~\mu$A\\
 $I_{-1}, I_{-2}$ & $I^c_{1a} - I^c_{1b}, I^c_{2a} - I^c_{2b}$ & $0~\mu$A\\
 $\beta_{1(2)}$ &  $2\pi L_{1(2)} I_{+1(2)} / \Phi_0$ & 1.35, 2.3\\
 $\delta \beta_{1(2)}$ &  $2\pi L_{1(2)} I_{-1(2)} / \Phi_0$ & 0 \\
 $\gamma_{1(2)}$ & $L_{1(2)}/2l_{1(2)}$  & 5, 9, 16 
\end{tabular}
\end{ruledtabular}
\end{subtable}

\begin{subtable}{(c) External parameters}
\centering
\begin{ruledtabular}
\begin{tabular}{l l}
\textrm{Symbol}&
\textrm{meaning}\\
\colrule
 $\px[1(2)]$ & Dimensionless flux threading through \\
 & the circuit loop containing $J_a$ and $L_1$ \\ & ($J_c$ and $L_2$) \\
 $\pxdc[1(2)]$ & Dimensionless flux threading through \\
 & the circuit loop containing $J_a$ and $J_b$ \\ & ($J_c$ and $J_d$)\\
 $m_{12}$ & Coupling between $L_1$ and $L_2$\\
\end{tabular}
\end{ruledtabular}  
\end{subtable}
\bigskip
\caption{Tables listing device fabrication parameters and calculated parameters and external parameters of the circuit. With $C = 1000$ and $L = 5~pH$, the time constant $t_c = 2.24~ps$. The values of $l_i$ are set at {0.16, 0.28, 0.5}pH, which correspond to $\gamma_i$ = {16, 9, 5}. The values of $I^c_{ia(b)}$ are set at {45, 75}$\mu A$, which correspond to $\beta_i$ = {1.35, 2.3}. The temperature is set at $4.2~K$, the normal boiling point of liquid helium.} 
\label{table:circuit_parameters}
\end{table}

\subsection{Work Done and Fidelity}

The work done by the $k$-th particle is expressed as:
\begin{align*}
W_k & = \sum^{n}_{i = 0} [U(\vec{x}_k(\tau_i), \tau_{i+1}) - U(\vec{x}_k(\tau_{i}), \tau_{i}) ]\\
  & ~~ -  \sum^{n}_{i = 0} [U_{\text{min}}(\tau_{i+1)} - U_{\text{min}}(\tau_{i)}]
  ~.
\end{align*}
Here, $\tau_i$ is the time at the $i$-th step, $\vec{x}_k$ is the state of the $k$-th particle, and $U$ is the potential, which is a function of the particle state and time. The upper limit $n$ of the summation is the total number of time steps.

The first summation term in the work done is the difference between the potential energy at $\tau_{i+1}$ of the $i$-th state and that at $\tau_{i+1}$ of the same state. The second summation is the difference of the minimum point of the potential at $\tau_{i+1}$ and $\tau_i$. This term is used to eliminate the work cost for shifting the potential up and down, which costs zero net work after a complete cycle. The average work done is the average of the work done by all the particles in the ensemble.

Fidelity is defined as the ratio of particles that successfully reach their intended final well to the total number of particles in the ensemble. It serves as a direct measure of a protocol's performance in guiding particles to their correct logical state. The truth tables for NAND, CE, and EF are provided in Figure \ref{fig:truth_table_and_three_operations}. Any particle whose final position is inconsistent with its corresponding truth table entry is classified as an error.

\begin{figure*}
\includegraphics[scale=0.42]{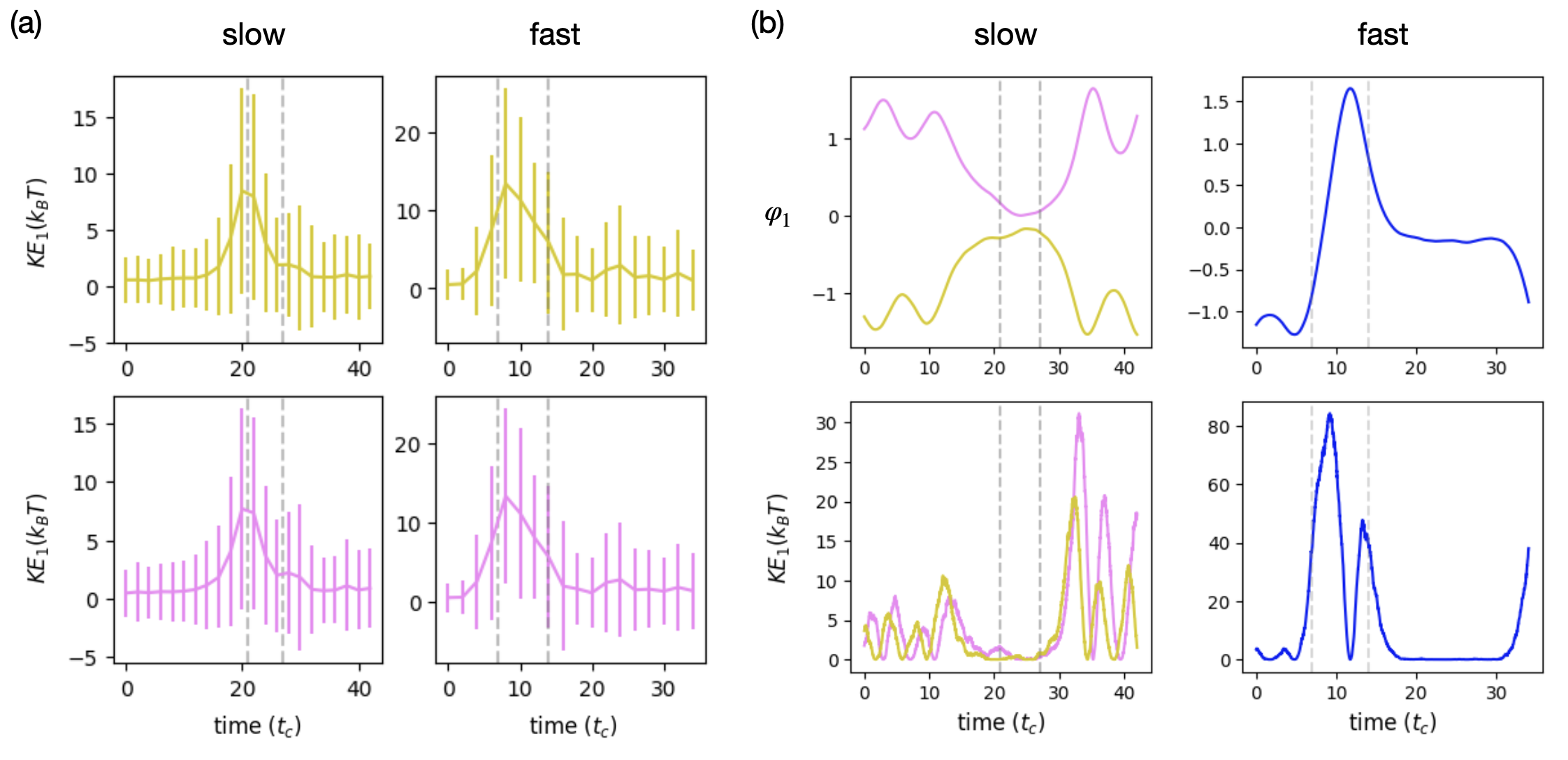}
\caption{(a) Average kinetic energy of the 01 (yellow) and 11 (pink) particles as a function of time for EF (slow) and EF (fast): In the slower protocol, some particles have near-zero kinetic energy during the barrier-raising substage (region between the grey dashed lines). In contrast, in the faster version, the error bars no longer reach 0, indicating that the particles of the storage pair maintain a nonzero speed and can successfully transition to the target well. (b) Example trajectories of failure-prone particles in both the slow and fast versions of the EF protocol. In the slow version, the primary source of error is the storage pair. During the barrier-raising substage, these particles have near-zero kinetic energy and remain very close to the boundary, preventing them from successfully reaching their target wells. In EF (fast), the error comes from the erasure pair instead because too much kinetic energy is input to the system. The graph shows a failure trajectory from a $00$ particle. (Animation: \textcolor{blue}{ \href{https://drive.google.com/file/d/1FHn_Dgznadt5yDeFAXeUzzZuqcRj5Ree/view?usp=sharing}{slow protocol}}, \textcolor{blue}{ \href{https://drive.google.com/file/d/1Rz7D6tsMzH3seL5_3_bQwmNwIqmDkTyG/view?usp=sharing}{fast protocol}}) }
\label{fig:EF_fidelity_improvement}
\end{figure*}

\subsection{dt analysis}

\begin{figure}
\includegraphics[scale=0.55]{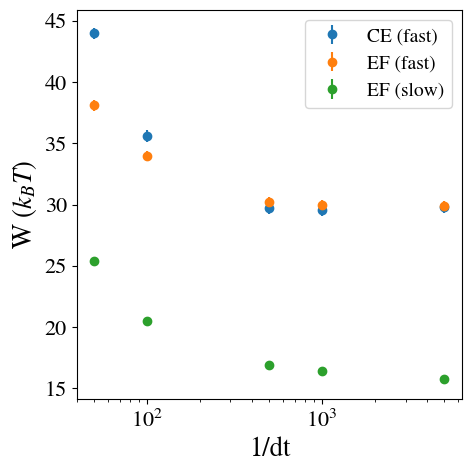}
\caption{The graph plots the converged work cost for the three protocols (CE (fast), EF (slow), and EF (fast)) vs $1/dt$, where $dt  = \{1/50, 1/100, 1/500, 1/1000, 1/5000\}$.}
\label{fig:dt_analysis}
\end{figure}

Figure \ref{fig:dt_analysis} investigates the relationship between the numerical time step (dt) and the mean work cost. This analysis considers dt values across the range of $[1/50, 1/100, 1/500, 1/1000, 1/5000]$ with $10,000$ trajectories. Based on this investigation, we selected $dt = 1/1000$ for all subsequent simulations because it provides the optimal balance between numerical accuracy and computational efficiency.

\subsection{Integration method}
A $4$th-order Runge-Kutta method was applied to the deterministic portion and the Euler-Maruyama method was used for the stochastic portion of the integration. 

\section{Comparing trajectories of slow and fast EF}
\label{appendix:slow_EF_vs_fast_EF}

This section illustrates the crucial role of kinetic energy and momentum in determining the fidelity of EF protocol. Figure \ref{fig:EF_fidelity_improvement}(a) plots the average kinetic energy for the $10$ and $11$ particles in both the EF (slow) and EF (fast) protocols. In the EF (slow) protocol, during the critical barrier-raising substage (the region between the grey dashed lines), the error bars scatter across zero. This indicates that some particles may fail to complete the required logical flip due to lacking KE. The first column of Figure \ref{fig:EF_fidelity_improvement}(b) illustrates a sample failure trajectory: during the barrier-raising substage, the failing particles remain clustered around $\p[1] = 0$. Since they do not acquire the necessary kinetic energy, they fail to complete their transition and consequently revert to their original wells.

The EF (fast) protocol successfully addresses this deficiency. Its average KE graphs show that the error bars are consistently far from zero during the barrier-raising substage, ensuring the $10$ and $11$ particles maintain high momentum and successfully transition into their target well. However, this substantial increase in speed simultaneously imparts excess kinetic energy to the $00$ particles. This excess energy causes error in the erasure pair. The second column of Figure \ref{fig:EF_fidelity_improvement}(b) illustrates a sample failure trajectory for EF (fast): the $00$ particle, having gained too much kinetic energy, fails to remain settled and reverts to its initial well, thereby causing a fidelity error in the erasure pair.

\section{Protocol Detail}
\label{appendix:detail_of_protocols}

\subsection{Work Cost of Conditional Tilt}

\begin{figure}
\includegraphics[scale=0.65]{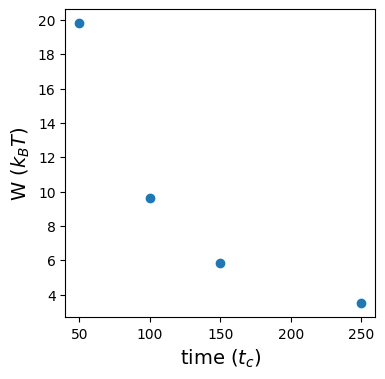}
\caption{Work cost of conditional tilt for different durations}
\label{fig:work_cost_of_CT}
\end{figure}

The protocol for Conditional Tilt (CT) is defined by Substages 4 through 6 in Figure \ref{table:protocol_table}(d) and is assigned a base duration of $50t_c$. To analyze the work-cost-versus-speed trade-off of the CT itself, we simulated its performance using the same protocol but extending the duration by factors of two, three, and five times the base duration. The resulting work cost of the conditional tilt as a function of these different time lengths is presented in Figure \ref{fig:work_cost_of_CT}. As expected theoretically, the work cost decreases as the duration increases. It would eventually approach Landauer's bound, but that may take an arbitrarily long protocol. The range displayed shows a reasonable swath of practical finite-time protocols with a reasonable balance of work cost and operating speed. There is not much energetic advantage in going beyond $250~t_c$, when compared to the extra time spent processing. Conversely, protocols that are shorter than $50~t_c$ become more and more costly and quickly dominate the work cost of the whole protocol.

\subsection{Protocols}

\begin{figure*}
\includegraphics[scale=0.7]{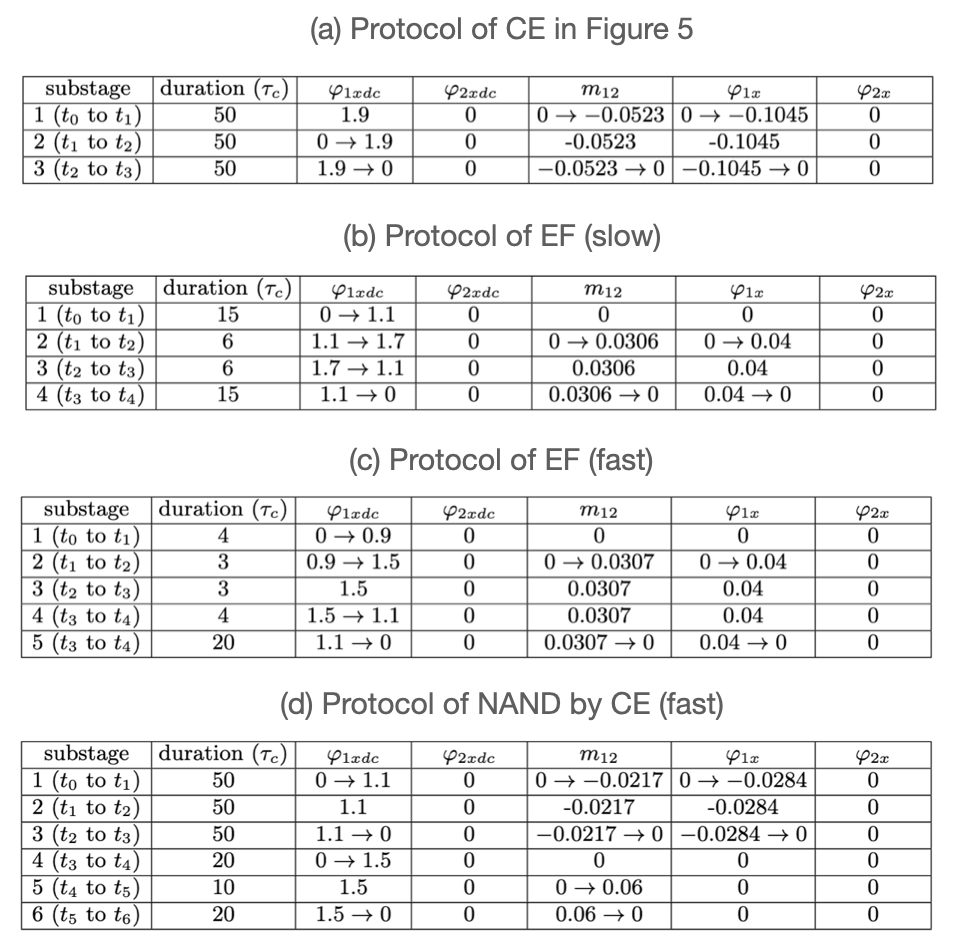}
\caption{The tables detail the specific time-dependent changes applied to the external control parameters for each simulation protocol: (a) Protocol of CE in Figure \ref{fig:work_distribution_for_CE}, (b) Protocol of EF (slow) in Figure \ref{fig:work_distribution_for_EF_1}(a), (c) Protocol of EF (fast) in Figure \ref{fig:work_distribution_for_EF_1}(b) and (d) Protocol of NAND by CE (fast) in Figure \ref{fig:work_distribution_of_NAND_by_CE_and_EF}.}
\label{table:protocol_table}
\end{figure*}

Figure \ref{table:protocol_table} presents the specific protocol tables used in the simulations. Figure \ref{table:protocol_table}(a) defines the CE protocol used in the simulation of Figure \ref{fig:work_distribution_for_CE}, which employed the parameters $\beta = 2.3$ and $\gamma = 9$. Figures \ref{table:protocol_table}(b), (c), and (d) define the protocols for EF (slow), EF (fast) and CE (fast) protocol, respectively. $\beta = 1.35$ in all the three cases, but $\gamma = 9$ for EF (slow), $\gamma = 5$ for EF (fast), and $\gamma = 16$ for CE (fast). All of the above simulations use the same capacitance, inductance, resistance, and critical currents as shown in Table \ref{table:circuit_parameters}.

\begin{figure*}
\includegraphics[scale=0.35]{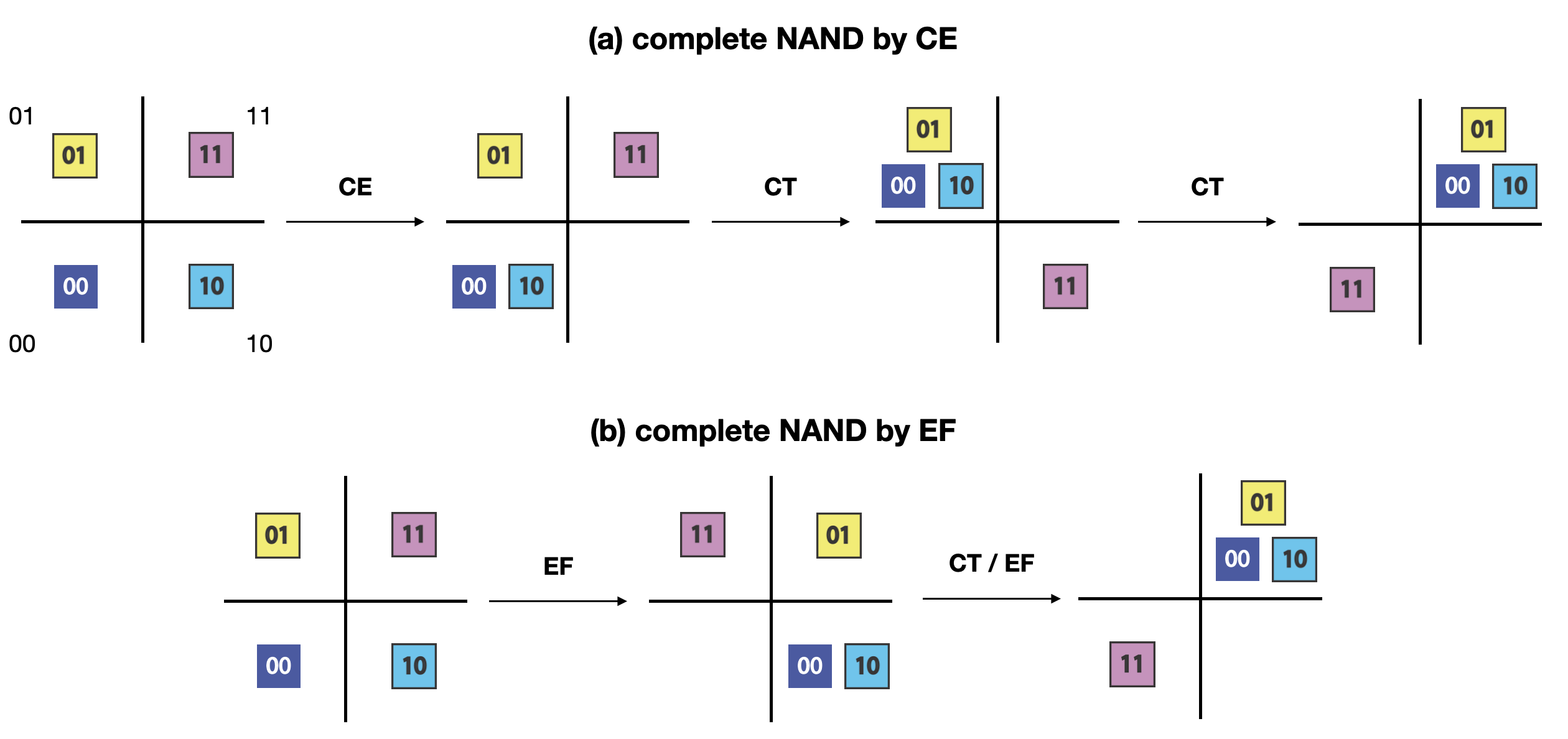}
\caption{Schematic illustration showing the complete NAND protocol by (a) CE and (b) EF.}
\label{fig:complete_NAND_by_CE_and_EF}
\end{figure*}

\begin{figure*}
\includegraphics[scale=0.48]{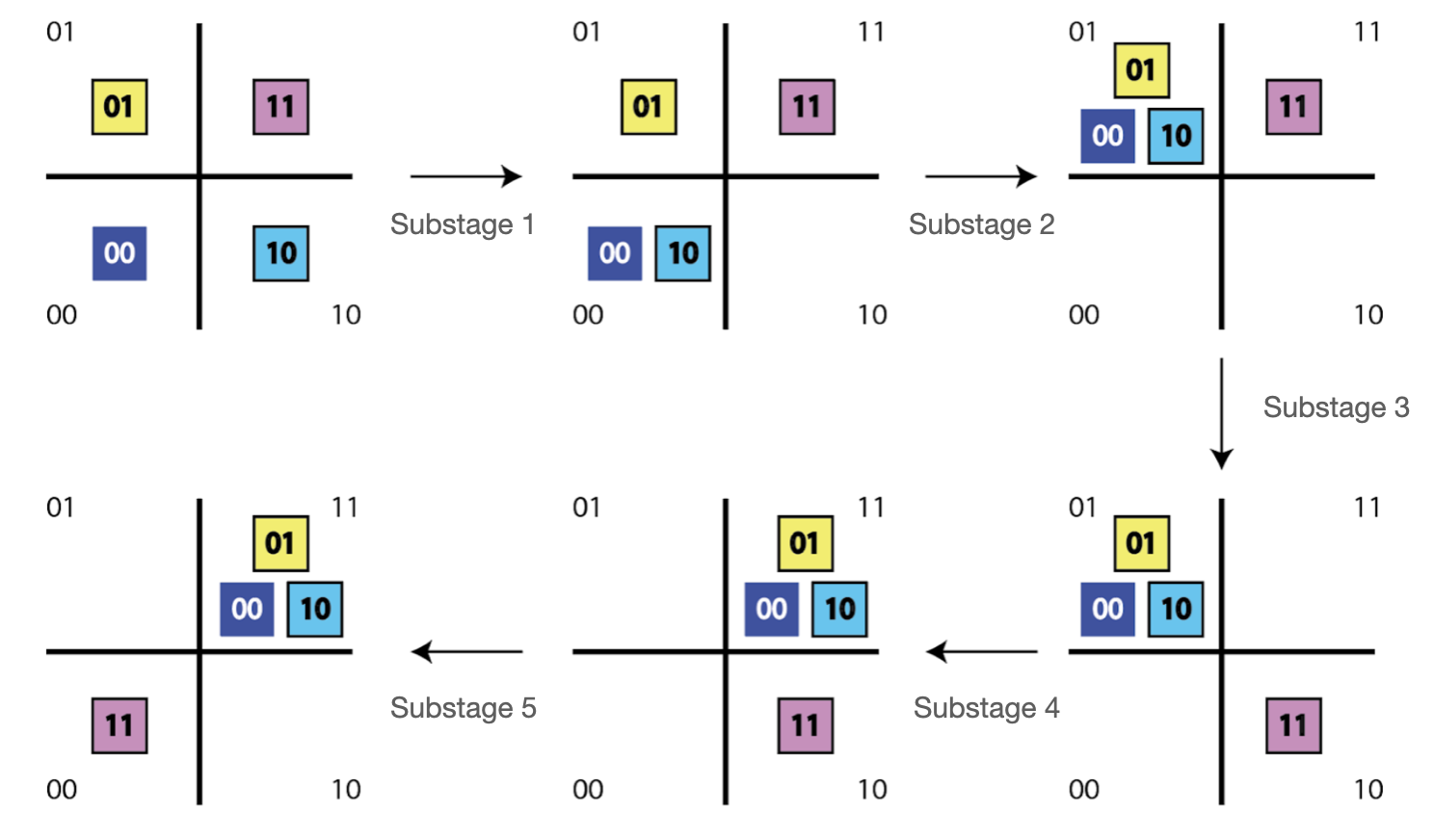}
\caption{Performing controlled erasure five times for complete NAND operation.}
\label{fig:longer_complete_erasure}
\end{figure*}

\section{Alternative Protocols for NAND}
\label{appendix:alternative_NAND}

\subsection{Complete NAND}

The analysis in the main text describes and presents simulation results for the partial NAND operation, implemented using both the CE and EF protocols. In this setting, the computation designates one bit as the primary output and the other as an irrelevant (garbage) bit. To eliminate the presence of the garbage bit and produce a clean, single-state result, one implements a complete NAND operation, which actively forces the logical states of the two output bits to be identical.

Figure \ref{fig:complete_NAND_by_CE_and_EF} illustrates the conceptual implementation of the complete NAND using both CE and EF. For the CE-based implementation, the complete NAND is achieved by appending one additional CT substage to the partial NAND protocol. This final manipulation moves $00$, $01$, and $10$ particles into the well $11$, while moving the particle $11$ into $00$ well at the end of the operation.

For the partial NAND by EF, the complete NAND can be realized by applying either one additional CT or one additional EF substage. This final step moves the $11$ particles to the $00$ well, while concurrently steering the other three particle types ($00$, $01$, and $10$) to the $11$ well, thereby ensuring both bits to be identical at the final state.

\subsection{Complete NAND by CE only}

An alternative complete NAND operation based off of CE only is described in Ref. \cite{pratt2024}. As illustrated in Figure \ref{fig:longer_complete_erasure}, this method involves performing CE protocol five times. The work cost for complete NAND by CE only (Figure \ref{fig:longer_complete_erasure}) can be estimated by observing which particles types must move from a higher well to a lower well in each subprocess. The following table summarizes the particle types that must move from a higher well into a lower well. By summing the work cost column, the work cost is $8$ times of the work cost of a single CE.

\begin{table}[h]
    \centering
    \begin{tabular}{ | wc{2cm} | wc{3cm} | wc{2cm}|} 
         \hline
         substage & type of particles & work cost\\
         \hline
         1 & 10  & 1   \\
         \hline
         2 & 00, 10  & 2   \\
         \hline
         3 & 11  & 1   \\
         \hline
         4 & 00, 01, 10  & 3   \\
         \hline
         5 & 11  & 1   \\
         \hline
    \end{tabular}
    \label{tab:placeholder}
\end{table}

\subsection{Partial NAND by CE only}

An alternative implementation of a partial NAND using three repeated CEs operations was also proposed in Ref. \cite{pratt2024}. As it requires three serial CE operations the duration is three times that of a single CE. The error rate and work costs differ, however, as they were calculated assuming a uniform starting distribution over the four wells. When taking into account the change of distribution for each subsequent operation, we still expect the error rate to be triple that of the single CE performed on a uniform distribution. However, the work cost for this implementation is four times that of the single CE---a manifestation of the additional work costs associated with modularity and variations in initial distribution \cite{boyd2018modularity,wolpert2020circuits}. In short, the CE-based NAND operation using a CT is simultaneously faster and more accurate, while also costing less work than one built completely out of CEs.

While these protocols incur higher work costs, lower fidelity, and longer durations, they may offer advantages in control complexity by exclusively using CE and forgoing the need for conditional tilts.

\section{Entropy change of partial NAND and complete NAND}
\label{appendix:fundamental_work_cost}
The fundamental work cost of a protocol is calculated through comparing the Shannon entropy difference between the initial and final distribution. This appendix shows how to calculate the fundamental work cost for a partial NAND and a complete NAND. We start with a four-well potential and the Shannon entropy of the initial distribution is $-\frac{1}{4} \ln\frac{1}{4} \times 4 = 2\ln2$.

The distribution after the first substage in Figure \ref{fig:complete_NAND_by_CE_and_EF}(b) corresponds to the final state of a partial NAND operation. In this state, two input states are mapped into a single well (total probability $1/2$), while the other two states remain separate (each with probability $1/4$). The Shannon entropy of this final distribution is:
$S_{\text{Partial}} = -\left( \frac{1}{2} \ln \frac{1}{2} + \frac{1}{4} \ln \frac{1}{4} + \frac{1}{4} \ln \frac{1}{4} \right) = \frac{3}{2} \ln 2$. The minimum work cost for the partial NAND is calculated from the entropy change $\Delta S$, which is about $0.347$.

On the other hand, the distribution after the second substage in Figure \ref{fig:complete_NAND_by_CE_and_EF}(b) corresponds to the final state of a complete NAND operation, where three out of four inputs are mapped to the logical $0$ state (total probability $3/4$), and one input remains in the logical $1$ state (probability $1/4$). The Shannon entropy of this final distribution is: $S_{\text{Complete}} = -\left( \frac{3}{4} \ln \frac{3}{4} + \frac{1}{4} \ln \frac{1}{4} \right) =  2 \ln 2 - \frac{3}{4} \ln 3$. The minimum work cost for the complete NAND is about $0.824$.

\end{document}